\documentclass[aps,prb,groupedaddress,amssym,floatfix,notitlepage,eqsecnum]{revtex4-1}

\usepackage{amsfonts}
\usepackage{amsmath}
\usepackage{amssymb}
\usepackage{graphicx}
\begin{document}


\title{Impurity entropy of junctions of multiple quantum wires}
\author{Zheng Shi}
\affiliation{Department of Physics and Astronomy, University of British Columbia,
Vancouver, BC, Canada V6T 1Z1}





\date{\today}

\begin{abstract}
We calculate the zero-temperature impurity entropy of a junction of multiple quantum wires of interacting spinless fermions. Starting from a given single-particle S-matrix representing a fixed point of the renormalization group (RG) flows, we carry out fermionic perturbation theory in the bulk interactions, with the perturbation series summed in the random phase approximation (RPA). The results agree completely with boundary conformal field theory (BCFT) predictions of the ground state degeneracy, and also with known RG flows through the $g$-theorem.
\end{abstract}



\maketitle

\section{Introduction}

At the critical point of a one-dimensional quantum system with a boundary, it is well known that at low temperatures the logarithm of the partition function takes the form $\ln Z=\left( c\pi /6v\right) \left( L /\beta\right) +\ln g$, where $c$ is the central charge of the bulk conformal field theory, $v$ is the ``speed of light'' in the bulk, $\beta =1/T$ is the inverse temperature, and $L\gg v\beta$ is the length of the system.\cite{PhysRevLett.67.161,*PhysRevB.48.7297} The length-independent boundary term $S^{\text{imp}} =\ln g$ is experimentally accessible as a thermodynamic impurity entropy. Here the universal ``ground state degeneracy'' $g$ characterizes the conformally invariant boundary condition (CIBC), and always decreases along the direction of the renormalization group (RG) flow, as required by the ``$g$-theorem''.\cite{PhysRevLett.67.161,PhysRevLett.93.030402} This provides a consistency check on the RG phase diagram. $S^{\text{imp}}$ also contributes to the ground state entanglement entropy of the system,\cite{1742-5468-2004-06-P06002} which allows its numerical extraction by the density matrix renormalization group (DMRG) method.\cite{PhysRevLett.96.100603}

The boundary conformal field theory (BCFT) allows us to determine the ground state degeneracy $g_{A}$ associated with a given CIBC, $A$, up to a normalization factor.\cite{Cardy1989581} For a critical system with CIBCs $A$ and $B$ imposed at its two ends, we can interchange space and imaginary time to write its partition function as $Z_{AB}=\left\langle A \left| \exp \left( -L H_{P}/v\right) \right| B \right\rangle$, where $H_{P}$ is the Hamiltonian subject to periodic boundary conditions along the new spatial direction of length $v \beta$. We have reinterpreted the CIBCs as the ``initial'' and ``final'' boundary states $\left| A \right\rangle$ and $\left| B \right\rangle$. These boundary states can be constructed explicitly from the eigenstates of $H_{P}$. In the $L\gg v\beta$ limit, we identify the ground state degeneracy $g_{A}=\left\langle 0|A \right\rangle$ (and similarly for $B$). Here $\left| 0 \right\rangle$ is the ground state of $H_{P}$, with a phase chosen so that $g_{A}$ is positive definite. This is how the ground state degeneracy has been obtained for various RG fixed points in junctions of multiple Luttinger liquid quantum wires,\cite{Wong1994403,PhysRevB.89.045133,1742-5468-2006-02-P02008,PhysRevB.77.155422} a class of one-dimensional quantum impurity models extensively studied due to their importance in quantum circuits.\cite{PhysRevLett.68.1220,*PhysRevB.46.15233,PhysRevB.47.4631,*PhysRevB.47.3827,PhysRevLett.94.136405,1742-5468-2006-02-P02008,PhysRevB.86.075451}

On the other hand, some RG fixed points with highly nontrivial CIBCs have so far eluded the BCFT treatment. One prominent example is the $M$ fixed point in the three-lead junction (or ``Y-junction'') of spinless electrons.\cite{PhysRevLett.94.136405,1742-5468-2006-02-P02008,PhysRevB.66.165327,PhysRevB.88.075131} The corresponding CIBC is difficult to write down in the bosonization framework commonly used to treat Luttinger liquids, and the boundary state is so far unknown. However, in the non-interacting limit and the $Z_{3}$ symmetric case, the $M$ fixed point is described by single-particle S-matrices with the maximum transmission probability $8/9$ allowed by unitarity. It is thus natural to turn to an alternate description in terms of fermion fields, which diagonalizes the non-interacting part of the Hamiltonian using its single-particle S-matrix, and treats the electron-electron interaction in quantum wires as a perturbation. The S-matrix is then viewed as a coupling constant which renormalizes under fermionic RG, under the assumption that RG fixed points are characterized by certain single-particle S-matrices. \cite{PhysRevLett.71.3351,*PhysRevB.49.1966,PhysRevB.66.165327,PhysRevB.68.035421,0295-5075-82-2-27001,*PhysRevB.80.045109,*LithJPhys.52.2353,PhysRevLett.105.266404,PhysRevB.84.155426,PhysRevB.88.075131} This fermionic method successfully captures the presence of the $M$ fixed point, and agrees with bosonization predictions of the RG phase diagram at least when the interaction is not too strongly attractive. To our knowledge, however, it has not been used to analyze the impurity entropy and the ground state degeneracy at the RG fixed points.

In this paper, we calculate the zero-temperature impurity entropy of a generic junction of multiple quantum wires, which is at a fixed point with an off-resonance single-particle S-matrix. This is accomplished by performing the fermionic perturbation theory on the logarithm of the partition function, both at the first order\cite{PhysRevLett.71.3351,PhysRevB.66.165327} and in the random phase approximation (RPA) of the bulk interaction strength.\cite{SovPhysJETP.38.202,0295-5075-82-2-27001,PhysRevB.84.155426,PhysRevB.88.075131} In addition to recovering previous results at the RG fixed points well understood by bosonic BCFT, we find the impurity entropy at the fixed points only accessible in the fermionic approach, and in particular the $M$ fixed point. We further verify that our zero-temperature impurity entropy obeys the $g$-theorem when compared with the fermionic RG phase diagrams. It should be emphasized that our calculation is always done for a particular single-particle fixed point S-matrix-- it does not flow as the energy scale is lowered. Unless the system is non-interacting, most S-matrices do not correspond to any fixed point for a given junction.

The outline of this paper is as follows. Section~\ref{sec:model} establishes a model for our junction. In Section~\ref{sec:perturb}, we first find the impurity entropy as a function of the single-particle S-matrix and the interaction strengths to the first order in interaction, then extend this calculation to infinite order in interaction with the help of the RPA. Section~\ref{sec:gtheorem} is an application of these results to various fixed points of the 2-lead junction and the Y-junction; we confirm the agreement between our approach and the bosonic BCFT at fixed points predicted by both theories, as well as the validity of the $g$-theorem. Section~\ref{sec:openquestions} concludes the paper and discusses some open questions. Appendix~\ref{sec:applnZ} presents more technical details of the RPA calculation. Finally, Appendix~\ref{sec:appRGflow} is a summary of the fixed points and the RG flows of the 2-lead junction and the Y-junction in the fermionic approach, which will be needed in Section~\ref{sec:gtheorem}.

\section{Model\label{sec:model}}

In this section we explain our system of interest, depicted in Fig.~\ref{fig:sketch}. The junction is located at the origin $x=0$, and all $N$ semi-infinite uniform quantum wires that it connects are aligned with the $+x$ axis. The Hamiltonian consists of three parts:

\begin{figure}
\includegraphics[width=0.6\textwidth]{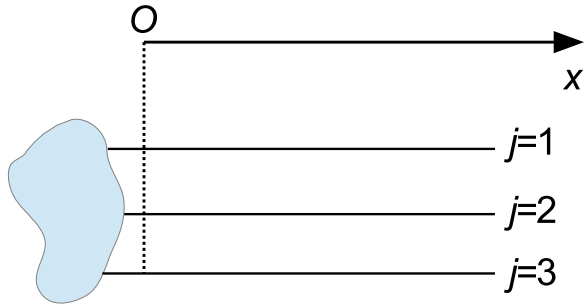}%
\caption{Sketch of a junction of 3 quantum wires.\label{fig:sketch}}
\end{figure}

\begin{equation}
H=\sum_{j=1}^{N}\left(  H_{0\text{,wire}}^{j}+H_{\text{int}}^{j}\right)
+H_{0,B}\text{.}%
\end{equation}
The bulk term $H_{0\text{,wire}}^{j}$ and the boundary term $H_{0,B}$ are quadratic in electron operators, and describe wire $j$ and the junction, respectively. $H_{\text{int}}^{j}$ is quartic and accounts for electron-electron interaction in the bulk of wire $j$.

Going to the continuum limit, we keep only the degrees of freedom in a narrow energy band of width $2D$ around the Fermi level:

\begin{equation}
\psi_{j}\left(  x\right)  \approx e^{ik_{Fj}x}\psi_{jR}\left(  x\right)
+e^{-ik_{Fj}x}\psi_{jL}\left(  x\right)  =\int_{-D}^{D}\frac{dE}{\sqrt{2\pi
v_{Fj}}}\left[  \psi_{jR}\left(  E\right)  e^{i\left(  \frac{E}{v_{Fj}}%
+k_{Fj}\right)  x}+\psi_{jL}\left(  E\right)  e^{-i\left(  \frac{E}{v_{Fj}%
}+k_{Fj}\right)  x}\right]  \text{.} \label{leftright}%
\end{equation}
Here $\psi_{j}\left(  x\right)$ annihilates an electron on wire $j$ at coordinate $x$. $v_{Fj}$ and $k_{Fj}\gg D/v_{Fj}$ are respectively the Fermi velocity and Fermi momentum in wire $j$ in the absence of the interaction, and the dispersion relation is assumed to be linear $E=E_{j}\left(  k\right)  =v_{Fj}k$. In terms of left-movers $\psi_{jL}$ and right-movers $\psi_{jR}$, the bulk Hamiltonian of wire $j$ is given by

\begin{equation}
H_{0\text{,wire}}^{j}\approx iv_{Fj}\int_{0}^{\infty}dx\left[  \psi_{jL}%
^{\dag}\partial_{x}\psi_{jL}-\psi_{jR}^{\dag}\partial_{x}\psi_{jR}\right]
\left(  x\right)  \approx\int_{-D}^{D}dE\,E\left(  \psi_{jR}^{\dag}\left(
E\right)  \psi_{jR}\left(  E\right)  -\psi_{jL}^{\dag}\left(  E\right)
\psi_{jL}\left(  E\right)  \right)  \text{,} \label{bulkquad}%
\end{equation}
and

\begin{equation}
H_{\text{int}}^{j}=\int_{0}^{\infty}dxg_{2}^{j} \psi
_{jR}^{\dag}\left(  x\right)  \psi_{jR}\left(  x\right)  \psi_{jL}^{\dag
}\left(  x\right)  \psi_{jL}\left(  x\right)  \text{.} \label{interaction0}%
\end{equation}

We have assumed short-range interaction, ignored Umklapp processes and retained only the $g_{2}$ processes of interactions between chiral densities of opposite chiralities.\cite{solyom2010fundamentals3} Interactions between chiral densities of the same chirality ($\psi_{R}^{\dag}\psi_{R}\psi_{R}^{\dag}\psi_{R}$ and $\psi_{L}^{\dag}\psi_{L}\psi_{L}^{\dag}\psi_{L}$) are also ignored; in the absence of $g_{2}$ processes they renormalize the Fermi velocity but not the Luttinger parameter.

The boundary term $H_{0,B}$ causes scattering at the junction and turns left-movers into right-movers. Instead of writing it down directly, we now introduce the scattering basis operators $\phi$ which diagonalize the full non-interacting Hamiltonian, $\sum_{j=1}^{N} H_{0\text{,wire}}^{j}+H_{0,B}$. This scattering basis transformation also reveals the boundary conditions satisfied by the left- and right-movers in terms of the single-particle S-matrix. We make the important assumption that the S-matrix elements at low energies at the fixed point in question are independent of the electronic energy $E$, $S_{jj^{\prime}}\left(  E\right)  \equiv S_{jj^{\prime}}$; in other words the junction is assumed to be off resonance. The scattering state of an incident electron from wire $j^{\prime}$ at energy $E^{\prime}$ then reads%

\begin{equation}
\phi_{j^{\prime}}^{\dag}\left(  E^{\prime}\right)  \left\vert 0\right\rangle
=\sum_{j}\int_{0}^{\infty}dx\frac{1}{\sqrt{2\pi v_{Fj}}}\left(  \delta
_{jj^{\prime}}e^{-i\frac{E^{\prime}}{v_{Fj}}x}\psi_{jL}^{\dag}\left(
x\right)  +S_{jj^{\prime}}e^{i\frac{E^{\prime}}{v_{Fj}}x}\psi_{jR}^{\dag
}\left(  x\right)  \right)  \left\vert 0\right\rangle +\ldots\text{,}
\label{scatstate}%
\end{equation}
where $\left\vert 0\right\rangle $ is the Dirac sea of left- and right-movers. This is an asymptotic expression valid well away from the junction. Eq.~(\ref{scatstate}) relates the original electron operators $\psi$ in terms of the scattering basis operators $\phi$,

\begin{subequations}
\label{scatbas}%
\begin{align}
\psi_{jR}\left(  E\right)   &  =\sum_{j^{\prime}=1}^{N}\int dE^{\prime}%
\int_{0}^{\infty}dx\left(  \frac{1}{\sqrt{2\pi v_{Fj}}}e^{i\frac{E}{v_{Fj}}%
x}\right)  ^{\ast}\left(  \frac{1}{\sqrt{2\pi v_{Fj}}}S_{jj^{\prime}}%
e^{i\frac{E^{\prime}}{v_{Fj}}x}\right)  \phi_{j^{\prime}}\left(  E^{\prime
}\right) \nonumber\\
&  =\sum_{j^{\prime}}\int_{-D}^{D}\frac{dE^{\prime}}{2\pi}\frac{-i}%
{E-E^{\prime}-i0}S_{jj^{\prime}}\phi_{j^{\prime}}\left(  E^{\prime}\right) \text{,}
\end{align}
and similarly

\begin{equation}
\psi_{jL}\left(  E\right)  =\int_{-D}^{D}\frac{dE^{\prime}%
}{2\pi}\frac{i}{E-E^{\prime}+i0}\phi_{j}\left(  E^{\prime}\right) \text{.}
\end{equation}
\end{subequations}

The quadratic Hamiltonian is diagonal in the scattering basis:

\begin{equation}
\sum_{j=1}^{N}H_{0\text{,wire}}^{j}+H_{0,B}=\sum_{j}\int dE\,E\phi_{j}^{\dag
}\left(  E\right)  \phi_{j}\left(  E\right) \text{.}  \label{quadratic}%
\end{equation}
Inserting the scattering basis transformation into Eq.~(\ref{interaction0}) we find the interaction in the scattering basis,

\begin{equation}
H_{\text{int}}^{j}=g_{2}^{j}\int_{0}^{\infty}dx \sum
_{l_{1}l_{2}l_{3}l_{4}}\int\frac{dE_{1}dE_{2}dE_{3}dE_{4}}{\left(
2\pi\right)  ^{2}v_{Fj}^{2}}\varrho_{l_{1}l_{2}l_{3}l_{4}}^{j}\left(
E_{1},E_{2},E_{3},E_{4};x\right)  \phi_{l_{1}}^{\dag}\left(  E_{1}\right)
\phi_{l_{2}}\left(  E_{2}\right)  \phi_{l_{3}}^{\dagger}\left(  E_{3}\right)
\phi_{l_{4}}\left(  E_{4}\right)  \text{,} \label{interaction}%
\end{equation}
where we introduce the function%

\begin{equation}
\varrho_{l_{1}l_{2}l_{3}l_{4}}^{j}\left(  E_{1},E_{2},E_{3},E_{4};x\right)
\equiv\frac{1}{2}\left[  e^{i\left(  -E_{1}+E_{2}+E_{3}-E_{4}\right)  \frac
{x}{v_{Fj}}}S_{jl_{1}}^{\ast}S_{jl_{2}}\delta_{jl_{3}}\delta_{jl_{4}%
}+e^{i\left(  -E_{3}+E_{4}+E_{1}-E_{2}\right)  \frac{x}{v_{Fj}}}S_{jl_{3}%
}^{\ast}S_{jl_{4}}\delta_{jl_{1}}\delta_{jl_{2}}\right]  \text{.}%
\end{equation}
We have symmetrized the function $\varrho$ so that $\varrho_{l_{1}l_{2}l_{3}l_{4}}^{j}\left(  E_{1},E_{2},E_{3},E_{4};x\right) =\varrho_{l_{3}l_{4}l_{1}l_{2}}^{j}\left(  E_{3},E_{4},E_{1},E_{2};x\right)$.

\section{Perturbation theory for the impurity entropy\label{sec:perturb}}

This section is devoted to a perturbation theory in interaction strength for the impurity entropy of our system.

Beginning from the zeroth order, we first argue that for a non-interacting system the ground state degeneracy must be the same for any single-particle S-matrix. Our argument goes as follows. Since the S-matrix is unitary, all its eigenvalues lie on the unit circle. Therefore, once we perform a unitary transformation which diagonalizes the S-matrix, any channel in the new basis only acquires a constant phase shift upon being scattered by the junction. (In the case of two-lead junctions the channels in the new basis are none other than the even and odd channels.)\cite{PhysRevB.82.165426} In the $L\to \infty$ limit, the low-energy finite-size spectrum of a system of $N$ decoupled wires is $N$-fold degenerate and evenly spaced. The effect of the junction is merely breaking the $N$-fold degeneracy by shifting each set of evenly spaced energy eigenvalues (corresponding to one of the new channels) by a constant amount, with the shift differing across all $N$ sets. At the partition function level this is indistinguishable from shifting the chemical potential for each new channel, which does not change the impurity entropy compared with the case of decoupled wires. This is consistent with the fact that in a non-interacting system any S-matrix corresponds to a fixed point, and any quadratic scattering at the junction is exactly marginal.

It should be mentioned that our argument is invalidated if states localized at the junction are coupled to the wires and lead to resonances close to the Fermi energy. In that case, the single-particle S-matrix and the phase shifts of every channel coupled to the localized state depend strongly on energy. In fact, Levinson's theorem requires that these phase shifts changes by $\pi$ as the energy is increased from well below to well above each resonance.\cite{Levinsontheorem,*1751-8121-41-29-295207} Consequently, the system picks up additional low-energy single-particle eigenstates in its finite-size spectrum. For every such additional eigenstate whose energy coincides with the Fermi energy in the thermodynamic limit, the impurity entropy increases by $\ln 2$ compared with the case of decoupled wires without the localized states. As stated earlier we shall ignore this scenario for simplicity, and focus on the off-resonance case.

\begin{figure}
\includegraphics[width=0.6\textwidth]{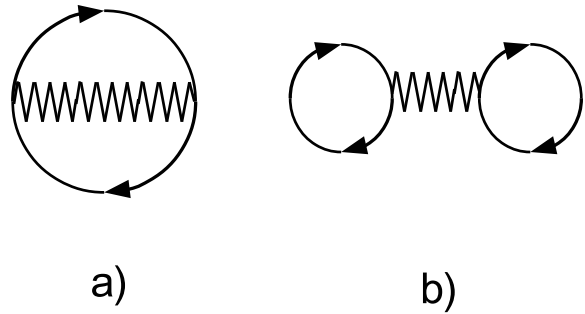}
\caption{The diagrams contributing to the logarithm of the partition function at the first order in interaction. The wavy line represents the interaction and the solid lines represent fermions in the scattering basis. The ``cracked egg'' diagram a) contributes to the impurity entropy, while the ``dumbbell'' diagram b) does not.\label{fig:OalphaIE}}%
\end{figure}

In the remainder of this section, we compute the logarithm of the partition function and extract the impurity contribution, first to the first order in $H_{\text{int}}^{j}$, then in the RPA. The perturbative corrections to the logarithm of the partition function are evaluated using the linked cluster expansion technique; for a review see Ref.~\onlinecite{mahan2000many}. At the first order in interaction, the correction is

\begin{equation}
\left(  \delta\ln Z\right)  ^{\left(  1\right)  }  =-\int d\tau\left\langle
T_{\tau}\sum_{j}H_{\text{int}}^{j}\left(  \tau\right)  \right\rangle \text{.}
\label{OalphalnZ}
\end{equation}
This is diagrammatically represented by Fig.~\ref{fig:OalphaIE}.

Inserting Eq.~(\ref{interaction}) into Eq.~(\ref{OalphalnZ}), Wick's theorem gives

\begin{align}
& \left\langle \phi _{l_{1}}^{\dag }\left( E_{1},\tau \right) \phi
_{l_{2}}\left( E_{2},\tau \right) \phi _{l_{3}}^{\dagger }\left( E_{3},\tau
\right) \phi _{l_{4}}\left( E_{4},\tau \right) \right\rangle   \notag \\
& =-\delta _{l_{1}l_{4}}\delta \left( E_{1}-E_{4}\right) \mathcal{G}%
_{l_{1}}\left( E_{1},0\right) \delta _{l_{2}l_{3}}\delta \left(
E_{2}-E_{3}\right) \mathcal{G}_{l_{2}}\left( E_{2},0\right)   \notag \\
& +\delta _{l_{1}l_{2}}\delta \left( E_{1}-E_{2}\right) \mathcal{G}%
_{l_{1}}\left( E_{1},0\right) \delta _{l_{3}l_{4}}\delta \left(
E_{3}-E_{4}\right) \mathcal{G}_{l_{3}}\left( E_{3},0\right)  \text{.}
\end{align}
Here $\mathcal{G}$ is the free scattering basis Matsubara Green's function, evaluated in the imaginary time domain with $\tau =0$. In the frequency space $\mathcal{G}_{j}\left(  E,i\omega_{n}\right)  =1/\left(  i\omega_{n}-E\right)  $, where $\omega_{n}=\left(  2n+1\right)  \pi/\beta$ is a fermion frequency. Replacing $\mathcal{G}_{l_{1}}\left( E_{1},\tau =0\right) =n_{F}\left( E_{1} \right)$, where $n_{F}\left(  \epsilon\right)  =1/\left(  e^{\beta\epsilon}+1\right)  $ is the Fermi distribution, we obtain

\begin{equation}
\left(  \delta\ln Z\right)  ^{\left(  1\right)  }  =-\beta\sum_{n}
\int_{0}^{\infty}dy\alpha_{n} \int
\frac{dE_{1}dE_{3}}{2\pi v_{Fn}}n_{F}\left(  E_{1}\right)  n_{F}\left(
E_{3}\right)  \left[  1-\left\vert S_{nn}\right\vert ^{2}\cos\left(  2\left(
E_{1}-E_{3}\right)  \frac{y}{v_{Fn}}\right)  \right]  \text{.}%
\label{OalphalnZ1}
\end{equation}
Here we have defined the dimensionless interaction strength,

\begin{equation}
\alpha_{j}=g_{2}^{j}/\left(  2\pi v_{Fj}\right)  \text{.} \label{dimlessinteraction}%
\end{equation}
The constant term in the square brackets, arising from the ``dumbbell'' diagram (Fig.~\ref{fig:OalphaIE}b), does not contribute to the impurity entropy. The reason is that it gives a term proportional to $\beta$, which corresponds to a shift of the ground state energy.\cite{1742-5468-2013-06-P06011} For the cosine term arising from the ``cracked egg'' diagram (Fig.~\ref{fig:OalphaIE}a), the $y$ integral yields a delta function. Using

\begin{align}
&  \int_{-D}^{D}dE_{1}dE_{3}n_{F}\left(  E_{1}\right)  n_{F}\left(
E_{3}\right)  \delta\left(  E_{1}-E_{3}\right) \nonumber\\
&  =\int_{-D}^{D}dE_{1}\frac{1}{\left(  e^{\beta E_{1}}+1\right)  ^{2}%
}=D+\frac{1}{\beta}\frac{1-e^{\beta D}}{1+e^{\beta D}}\text{,}%
\end{align}
we find in the $\beta D\gg1$ limit the part of $\left(  \delta\ln Z\right) ^{\left(  1\right)  }$ which is independent of $\beta$,

\begin{equation}
\left(  \delta\ln Z\right)  ^{\left(  1\right)  \text{,imp}}=-\frac{1}{4}
\sum_{n=1}^{N}\alpha_{n}\left\vert S_{nn}\right\vert ^{2}\text{.}%
\end{equation}
As the impurity entropy is defined up to an additive constant, it is convenient to benchmark it against the fixed point of all wires being decoupled from each other [known as the $N$ (von Neumann) fixed point in literature]. We thus have the first-order impurity entropy measured relative to the $N$ fixed point,

\begin{equation}
S^{\text{imp,}\left(  1\right)  }=\frac{1}{4}\sum_{n=1}^{N}\alpha_{n}\left(
1-W_{nn}\right)\text{.} \label{Oalphaimpent}
\end{equation}
where we define $W_{jj^{\prime}}\equiv\left\vert S_{jj^{\prime}}\right\vert ^{2}$ as the reflection/transmission probability matrix at the fixed point in question. Because $\left\vert S_{nn}\right\vert ^{2}\leq1$, this result is by virtue of the $g$-theorem consistent with the proposal that the $N$ fixed point is the most stable one for weak repulsive interaction in the wires ($\alpha_{n}>0$ for any $n$), and the most unstable one for weak attractive interaction in the wires ($\alpha_{n}<0$ for any $n$).\cite{PhysRevLett.71.3351,*PhysRevB.49.1966,PhysRevB.66.165327}

\begin{figure}
\includegraphics[width=0.7\textwidth]{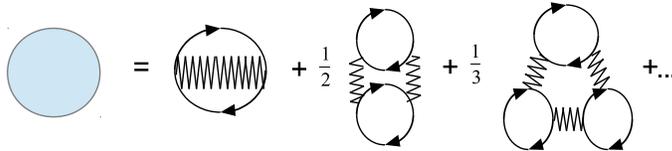}
\caption{The RPA diagrammatics of the impurity entropy in the RPA. The constant coefficients arise from linked cluster expansion.\label{fig:RPA}}
\end{figure}

We now compute the impurity entropy in the RPA. The linked cluster theorem states that the $m$th order correction to the logarithm of the partition function is

\begin{equation}
\left(  \delta\ln Z\right)  ^{\left(  m\right)  }=\frac{\left(  -1\right)
^{m}}{m}\int d\tau_{1}\cdots d\tau_{m}\left\langle T_{\tau}\sum_{n_{1}\cdots
n_{m}}H_{\text{int}}^{n_{1}}\left(  \tau_{1}\right)  \cdots H_{\text{int}%
}^{n_{m}}\left(  \tau_{m}\right)  \right\rangle _{\text{C}}\text{,}%
\label{lnZm}
\end{equation}
where the subscript C indicates that the imaginary time-ordered product contains different and connected diagrams only.\cite{mahan2000many} The RPA keeps a special subset of these diagrams, namely the ring diagrams (see Fig.~\ref{fig:RPA}). The details of this calculation are left for Appendix~\ref{sec:applnZ}; here we only present the final result. The impurity entropy relative to the $N$ fixed point is given by

\begin{equation}
S^{\text{imp,RPA}}=\frac{1}{2}\ln\det\left(  \mathbf{1}-\mathcal{Q}^{-1}W\right)
-\frac{1}{2}\sum_{n}\ln\frac{2K_{n}}{1+K_{n}}\text{,} \label{RPAimpent}%
\end{equation}
where $K_{j}=\sqrt{\left(  1-\alpha_{j}\right)  /\left(  1+\alpha_{j}\right)  }$ is the Luttinger parameter of wire $j$,\cite{solyom2010fundamentals3}, and

\begin{equation}
\mathcal{Q}_{jj^{\prime}}=Q_{j}\delta_{jj^{\prime}}\text{,}
Q_{j}=\frac{1+K_{j}}{1-K_{j}}\text{.} \label{scriptQ}%
\end{equation}

Eqs.~(\ref{Oalphaimpent}) and (\ref{RPAimpent}) are the central findings of this paper. They express the zero-temperature impurity entropy at a fixed point as a function of the corresponding non-interacting single-particle S-matrix and the interaction strengths in the wires.

\section{Application to 2-lead junctions and Y-junctions\label{sec:gtheorem}}

In this section we calculate the zero-temperature impurity entropy at known fixed points of 2-lead junctions and Y-junctions, both at the first order [Eq.~(\ref{Oalphaimpent})] and in the RPA [Eq.~(\ref{RPAimpent})]. The results are checked against BCFT predictions\cite{Wong1994403,PhysRevB.89.045133,1742-5468-2006-02-P02008} and tested for consistency with the $g$-theorem and the fermionic RG phase diagrams.\cite{PhysRevLett.71.3351,PhysRevB.66.165327,0295-5075-82-2-27001,PhysRevB.84.155426,PhysRevB.88.075131} Various fixed points are only briefly explained here; their properties are discussed in more detail in Appendix~\ref{sec:appRGflow}.

\subsection{2-lead junction}

Two fixed points exist in a 2-lead junction both at the first order and in the RPA [see Eqs.~(\ref{Oalpha2leadprob}) and (\ref{RPA2leadprob})].\cite{PhysRevLett.71.3351,0295-5075-82-2-27001} Apart from the complete reflection fixed point [i.e. the $N$ fixed point], there is also a perfect transmission fixed point [the $D$ (Dirichlet) fixed point] where backscattering at the junction is absent, $W_{11}=W_{22}=0$. Consequently, the zero-temperature impurity entropy of the $D$ fixed point reads at the first order

\begin{subequations}
\begin{equation}
S^{\text{imp,}D}=\frac{1}{4}\left(  \alpha_{1}+\alpha_{2}\right)  \text{,}
\end{equation}
and in the RPA

\begin{equation}
S^{\text{imp,}D}=\frac{1}{2}\ln \frac{2 K_{1} K_{2}}{K_{1}+K_{2}}  \text{.} \label{RPA2Dimpent}
\end{equation}
\end{subequations}

The RPA result agrees with the BCFT prediction,\cite{PhysRevB.89.045133} and both agree with the direction of the RG flow, which is from $D$ to $N$ when $\alpha_{1}+\alpha_{2}>0$ at the first order or $K_{1}^{-1}+K_{2}^{-1}>2$ in the RPA (see Appendix~\ref{sec:appRGflow}).\cite{LithJPhys.52.2353}

\subsection{Y-junction}

In the Y-junction, the $N$, $A_{j}$ and $\chi^{\pm}$ fixed points have reflection and transmission probabilities which are completely independent of the interaction strengths in the wires; hence they are known as ``geometrical'' fixed points. These fixed points are not only predicted by the fermionic RG [see Eqs.~(\ref{Oalpha3leadprob}) and (\ref{RPA3leadprob})]\cite{PhysRevB.66.165327,PhysRevB.84.155426,PhysRevB.88.075131} but also well understood through bosonization and BCFT.\cite{1742-5468-2006-02-P02008,PhysRevB.86.075451} In contrast, the ``non-geometrical'' fixed point $M$ (along with similarly non-geometrical $Q$ and $C^{\pm}$ in the RPA) is predicted by fermionic RG but not directly by bosonization or BCFT. We first discuss the impurity entropies at the geometrical fixed points in a generic asymmetric Y-junction, then move on to study the impurity entropies of non-geometrical fixed points in an asymmetric junction at the first order, and in a 1-2 symmetric junction in the RPA.

\subsubsection{Geometrical fixed points}

The non-trivial geometrical fixed points are the asymmetric fixed points $A_{j}$ and the chiral fixed points $\chi^{\pm}$. At $A_{3}$, wire 3 is decoupled from the wires 1 and 2 which are perfectly connected with each other, so that $W_{33}=W_{12}=W_{21}=1$ and the remaining $W$ elements vanish. Meanwhile, at $\chi^{\pm}$, an electron incident from wire $j$ is perfectly transmitted into wire $j\pm1$ (here $j+3\equiv j$); hence $W_{jj^{\prime}}=\left( 1-\delta_{jj}\mp\epsilon_{jj^{\prime}}\right)  /2$, where $\epsilon_{jj^{\prime}}$ is totally anti-symmetric and obeys $\epsilon_{12}=\epsilon_{23}=\epsilon_{31}=1$. According to Eq.~(\ref{Oalphaimpent}) the zero-temperature impurity entropy at $A_{3}$ and $D$ read at the first order

\begin{subequations}
\begin{equation}
S^{\text{imp,}A_{3}}=\frac{1}{4}\left(  \alpha_{1}+\alpha_{2}\right)  \text{,}
\end{equation}

\begin{equation}
S^{\text{imp,}\chi^{\pm}}=\frac{1}{4}\left(  \alpha_{1}+\alpha_{2}+\alpha_{3}\right)  \text{.}
\end{equation}
Noting that $S^{\text{imp,}A_{1}}-S^{\text{imp,}A_{3}}=\left(  \alpha_{3}-\alpha_{1}\right)  /4$ and $S^{\text{imp,}\chi^{\pm}}-S^{\text{imp,}A_{3}}=\alpha_{3}/4$, we infer from the $g$-theorem that $A_{3}$ should be more stable than $N$ if $\alpha_{1}+\alpha_{2}<0$, more stable than $A_{1}$ if $\alpha_{1}<\alpha_{3}$, and more stable than $\chi^{\pm}$ if $\alpha_{3}>0$; in addition, $\chi^{\pm}$ should be more stable than $N$ if $\alpha_{1}+\alpha_{2}+\alpha_{3}<0$. All these are in agreement with the local stability analysis in
Appendix~\ref{sec:appRGflow}.

In the RPA, from Eq.~(\ref{RPAimpent}),

\begin{equation}
S^{\text{imp,}A_{3}}=\frac{1}{2}\ln \frac{2 K_{1} K_{2}}{K_{1}+K_{2}}  \text{,}
\label{RPAA3impent}
\end{equation}

\begin{equation}
S^{\text{imp,}\chi^{\pm}}=\frac{1}{2}\ln \frac{K_{1}+K_{2}+K_{3}+K_{1}K_{2}K_{3}}{4K_{1}K_{2}K_{3}}  \text{.}
\label{RPAchiimpent}
\end{equation}
\end{subequations}
We observe that in the special case $K_{1}=K_{2}=K_{3}=K$, Eq.~(\ref{RPAchiimpent}) has been found in Ref.~\onlinecite{1742-5468-2006-02-P02008}.

Since wire $3$ is decoupled at both $A_{3}$ and $N$, it is not surprising that $S^{\text{imp,}A_{3}}$ is equal to $S^{\text{imp,}D}$ for the two-lead junction; by the $g$-theorem it is also consistent with the fact that $N$ is more stable than $A_{3}$ when $K_{1}^{-1}+K_{2}^{-1}>2$. Local stability analysis in Appendix~\ref{sec:appRGflow} indicates that when the RG flows are from $\chi^{\pm}$ to $N$, $K_{1}^{-1}+K_{2}^{-1}+K_{3}^{-1}>3$ and $4\left( K_{1}K_{2}+K_{2}K_{3}+K_{3}K_{1}\right)  <3\left(  K_{1}+K_{2}+K_{3}+K_{1}K_{2}K_{3}\right)  $; these conditions imply $S^{\text{imp,}\chi^{\pm}}>0$. On the other hand, when the RG flows are from $\chi^{\pm}$ to $A_{3}$, $K_{3}^{-1}+\left(  1+K_{1}K_{2}\right)  /\left(  K_{1}+K_{2}\right)  >2$, so $S^{\text{imp,}\chi^{\pm}}>S^{\text{imp,}A_{3}}$. In both cases we find agreement with the $g$-theorem.

\subsubsection{Non-geometrical fixed points}

Let us begin our study with a generic asymmetric Y-junction at the first order [see Eq.~(\ref{Oalpha3leadprob})], which features the prototype non-geometrical fixed point $M$. The transmission/reflection probabilities at $M$ are given by

\begin{equation}
W_{jj^{\prime}}=%
\genfrac{\{}{.}{0pt}{}{\left(  \frac{\alpha_{1}\alpha_{2}\alpha_{3}/\alpha
_{j}}{\alpha_{1}\alpha_{2}+\alpha_{2}\alpha_{3}+\alpha_{3}\alpha_{1}}\right)
^{2}\text{, }j=j^{\prime}}{\left(  1-\frac{\alpha_{1}\alpha_{2}\alpha
_{3}/\alpha_{j}}{\alpha_{1}\alpha_{2}+\alpha_{2}\alpha_{3}+\alpha_{3}%
\alpha_{1}}\right)  \left(  1-\frac{\alpha_{1}\alpha_{2}\alpha_{3}%
/\alpha_{j^{\prime}}}{\alpha_{1}\alpha_{2}+\alpha_{2}\alpha_{3}+\alpha
_{3}\alpha_{1}}\right)  \text{, }j\neq j^{\prime}}%
\text{.} \label{OalphaMW}%
\end{equation}
Of course, $M$ only exists when all matrix elements of $W$ as given in Eq.~(\ref{OalphaMW}) are between $0$ and $1$. For $Z_{3}$ symmetric interactions we simply have $W_{jj^{\prime}}=1/9+\delta_{jj^{\prime}}/3$. Substituting Eq.~(\ref{OalphaMW}) into Eq.~(\ref{Oalphaimpent}) we find

\begin{equation}
S^{\text{imp,}M}=\frac{\left(  \alpha_{1}+\alpha_{2}\right)  \left(
\alpha_{2}+\alpha_{3}\right)  \left(  \alpha_{3}+\alpha_{1}\right)  }{4\left(
\alpha_{1}\alpha_{2}+\alpha_{2}\alpha_{3}+\alpha_{3}\alpha_{1}\right)
}\text{.}%
\end{equation}
Since $S^{\text{imp,}M}-S^{\text{imp,}\chi^{\pm}}=\lambda_{M\text{,}1}/4$ [where $\lambda_{M\text{,}1}$ is given in Eq.~(\ref{OalphaMexps})], the $g$-theorem requires that $\chi^{\pm}$ should be more stable than $M$ if $\lambda_{M\text{,}1}>0$.

In the presence of 1-2 symmetric interactions ($\alpha_{1}=\alpha_{2}$), whenever the $M$ fixed point exists, we can show that

\begin{subequations}
\begin{equation}
\operatorname*{sgn}S^{\text{imp,}M}=\operatorname*{sgn}\frac{\left(
\alpha_{1}+\alpha_{3}\right)  ^{2}}{2\left(  \alpha_{1}+2\alpha_{3}\right)
}=\operatorname*{sgn}\alpha_{3}\text{,}
\end{equation}

\begin{equation}
\operatorname*{sgn}\left(  S^{\text{imp,}M}-S^{\text{imp,}A_{1}}\right)
=\operatorname*{sgn}\frac{\alpha_{1}\left(  \alpha_{1}+\alpha_{3}\right)
}{4\left(  \alpha_{1}+2\alpha_{3}\right)  }=\operatorname*{sgn}\alpha
_{1}\text{,}
\end{equation}

\begin{equation}
\operatorname*{sgn}\left(  S^{\text{imp,}M}-S^{\text{imp,}A_{3}}\right)
=\operatorname*{sgn}\frac{\alpha_{3}^{2}}{2\left(  \alpha_{1}+2\alpha
_{3}\right)  }=\operatorname*{sgn}\alpha_{3}\text{;}%
\end{equation}
\end{subequations}
thus the $g$-theorem dictates that $M$ should be more stable than $A_{1}$ and $A_{2}$ if $\alpha_{1}<0$, and more stable than $N$ and $A_{3}$ if $\alpha_{3}<0$. All of the above are again in accord with local stability analysis.

\begin{figure}
\includegraphics[width=0.4\textwidth]{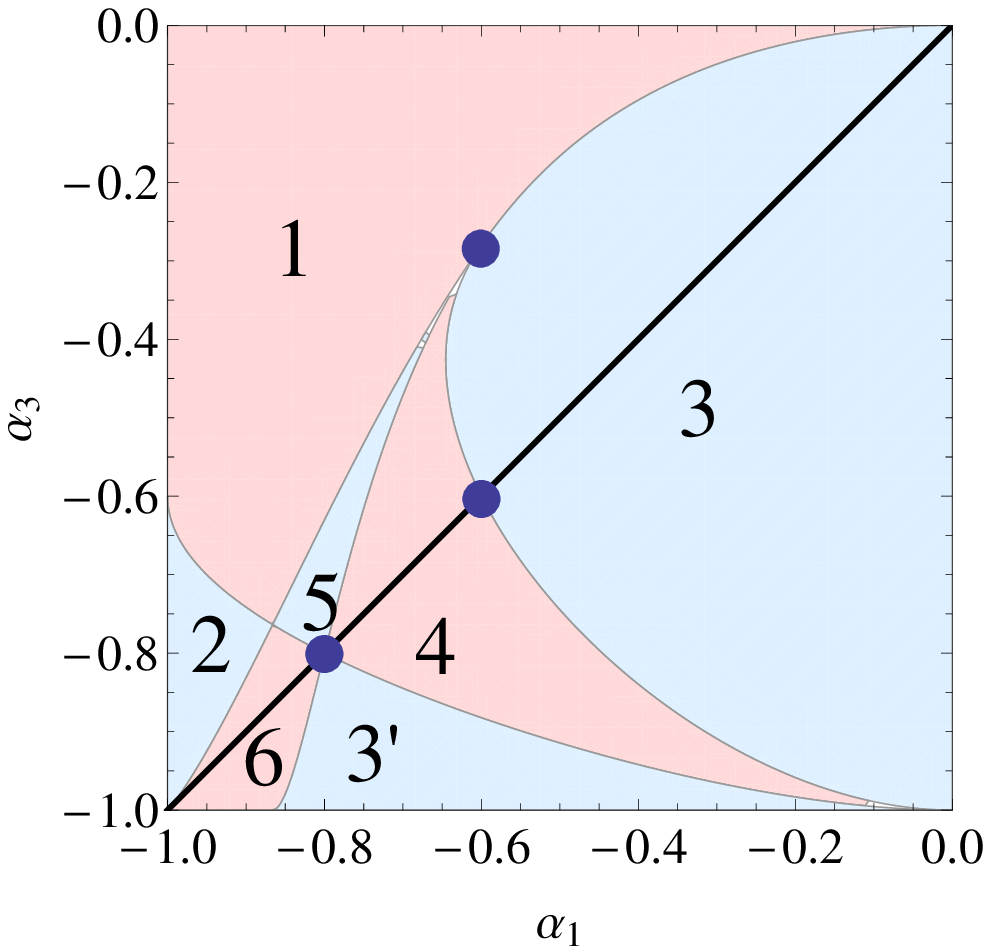}
\includegraphics[width=0.4\textwidth]{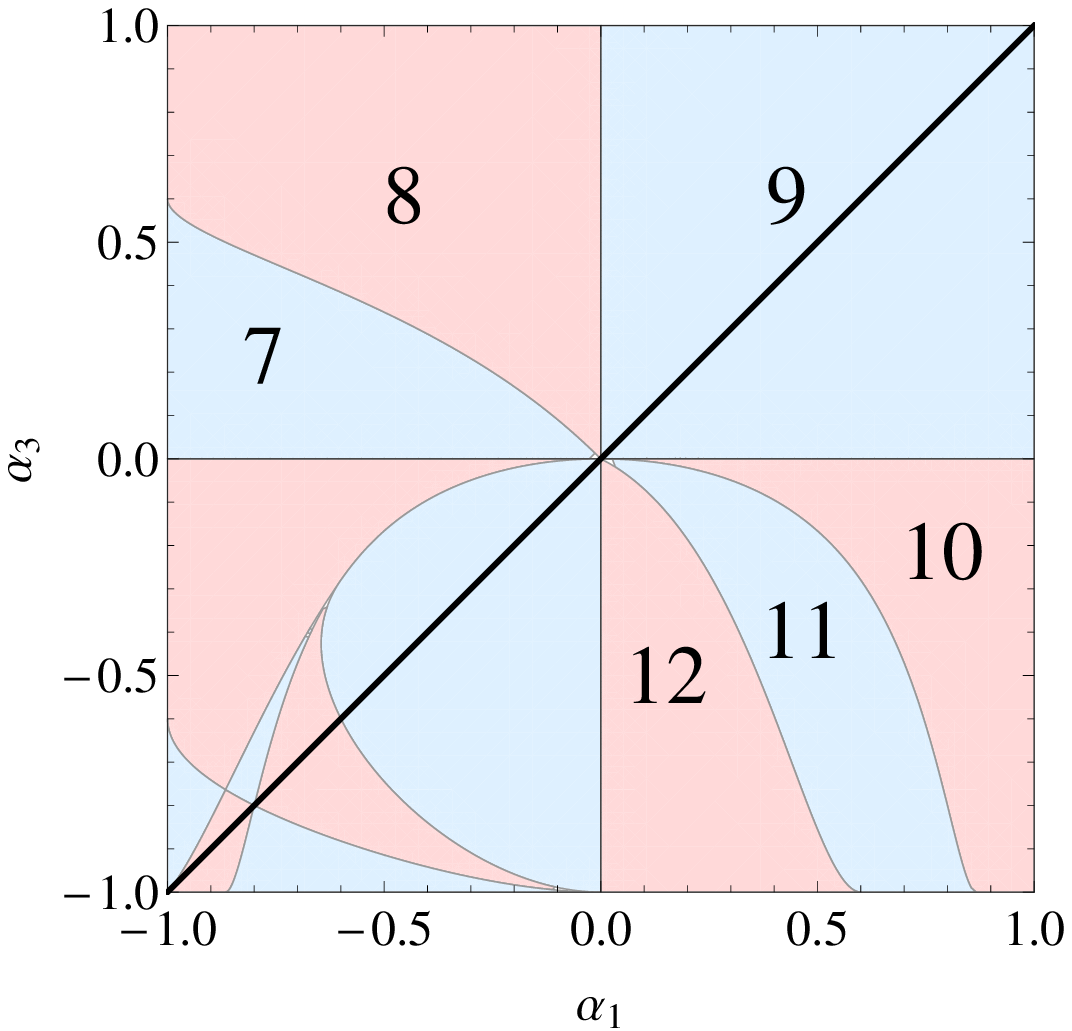}
\caption{The RPA RG phase portrait of the 1-2 asymmetric Y-junction found by Ref.~\onlinecite{PhysRevB.88.075131}. The regions 7-12 here are labeled as I-VI in Ref.~\onlinecite{PhysRevB.88.075131}. The first panel is a magnification of the region of attractive interactions from the second panel; the three dots there are the special points $K_{1}=K_{3}=2$, $K_{1}=K_{3}=3$ and $K_{1}=2$, $K_{3}=4/3$. A comparison of the impurity entropies of various fixed points on this phase portrait is given in Table~\ref{tab:RPA3leadFPIE}, and their local stability properties are described in Table~\ref{tab:RPA3leadFPstab}.\label{fig:3leadRPAphpo}}
\end{figure}

The generic $Z_{3}$ asymmetric Y-junction under the RPA is a very difficult problem [as evidenced by Eq.~(\ref{RPA3leadprob})], but if the junction is 1-2 symmetric, i.e. $K_{1}=K_{2}$, $W_{13}=W_{32}$ and $W_{23}=W_{31}$, then $A_{1}$ and $A_{2}$ cannot be reached by the RG flow, and the non-geometrical fixed points $M$, $Q$ and $C^{\pm}$ become analytically accessible.\cite{PhysRevB.84.155426,PhysRevB.88.075131} $M$ is a natural generalization of its namesake at the first order, while $Q$ and $C^{\pm}$ are always unstable and only appear for strong enough interactions: $Q$ is an asymmetric fixed point which may merge with $M$ under certain circumstances, and $C^{\pm}$ are a pair of chiral fixed points which may merge with $\chi^{\pm}$. Their properties are summarized in Appendix~\ref{sec:appRGflow}; here we merely list their impurity entropies and confirm that the $g$-theorem is obeyed. Inserting Eqs.~(\ref{abcMQ}) and (\ref{abcCpm}) into Eq.~(\ref{abccbar}) and then Eq.~(\ref{RPAimpent}), we find

\begin{subequations}
\begin{align}
& S^{\text{imp,}M\left(  Q\right)  }\nonumber\\
& =\frac{1}{2}\ln\left[  \frac{\left(  1+K_{1}\right)  \left(  K_{1}%
K_{3}+K_{3}-2\right)  }{9K_{3}\left(  K_{1}-1\right)  ^{2}}-\frac{\left(
3K_{1}K_{3}-K_{3}-2K_{1}\right)  \tau_{0}^{2}\left[  2\left(  1+K_{1}\right)
\pm\left\vert 1-K_{1}\right\vert \tau_{0}\right]  }{108K_{3}K_{1}^{2}}\right]
\text{,}%
\end{align}
where the plus sign is for $M$, and

\begin{equation}
S^{\text{imp,}C^{\pm}}=\frac{1}{2}\ln\frac{2K_{1}^{2}K_{3}\left(
K_{3}-1\right)  }{\left(  3K_{1}K_{3}-K_{3}-2K_{1}\right)  ^{2}}\text{.}%
\end{equation}
\end{subequations}
These allow us to compare the impurity entropies of various fixed points on the entire RG phase portrait,\cite{PhysRevB.88.075131} shown in Fig.~\ref{fig:3leadRPAphpo}. The results are listed in Table~\ref{tab:RPA3leadFPIE}. As expected, under the $g$-theorem they are fully consistent with the local stability analysis carried out in Table~\ref{tab:RPA3leadFPstab}.

\begin{table}
\caption{Comparison of the RPA impurity entropies of 1-2 symmetric fixed points based on the RG phase portrait on the $\alpha_{1}$-$\alpha_{3}$ plane, Fig.~\ref{fig:3leadRPAphpo}. $\left(  \chi^{\pm},M\right)  $ indicates that the relation between $S^{\text{imp,}\chi^{\pm}}$ and $S^{\text{imp,}M}$ is indeterminate in the corresponding region.\label{tab:RPA3leadFPIE}}
\begin{tabular}[c]{|l|l|}
\hline
Region & Impurity entropy $S^{\text{imp}}$\\ \hline
1 & $A_{3}<\chi^{\pm}<N$\\ \hline
2 & $A_{3}<C^{\pm}<\chi^{\pm}<N$\\ \hline
3 & $\chi^{\pm}<M<A_{3}<N$\\ \hline
4 & $\left(  \chi^{\pm},M\right)  <C^{\pm}<A_{3}<N$\\ \hline
5 & $\left(  A_{3},M\right)  <Q<\chi^{\pm}<N$\\ \hline
6 & $\left(  A_{3},M\right)  <Q<C^{\pm}<\chi^{\pm}<N$\\ \hline
3' & $M<\chi^{\pm}<A_{3}<N$\\ \hline
7 & $A_{3}<\chi^{\pm}<N$\\ \hline
8 & $A_{3}<\left(  N,\chi^{\pm}\right)  <M$\\ \hline
9 & $N<A_{3}<M<\chi^{\pm}$\\ \hline
10 & $N<A_{3}<M<\chi^{\pm}$\\ \hline
11 & $N<\chi^{\pm}<A_{3}$\\ \hline
12 & $M<\left(  N,\chi^{\pm}\right)  <A_{3}$\\ \hline
\end{tabular}
\end{table}

Let us now concentrate on the $Z_{3}$ symmetric Y-junction, $K_{1}=K_{2}=K_{3}=K$. Eq.~(\ref{RPAimpent}) gives the following impurity entropies at $\chi^{\pm}$, $M$, and $C^{\pm}$:%

\begin{subequations}
\begin{equation}
S^{\text{imp,}\chi^{\pm}}=\frac{1}{2}\ln\frac{3+K^{2}}{4K^{2}}\text{,}%
\end{equation}

\begin{equation}
S^{\text{imp,}M}=\frac{1}{2}\ln\frac{\left(  2+K\right)  ^{2}}{9K^{2}}\text{,}%
\label{RPAMimpent}
\end{equation}

\begin{equation}
S^{\text{imp,}C^{\pm}}=\frac{1}{2}\ln\frac{2K}{9\left(  K-1\right)  }\text{.}%
\end{equation}
\end{subequations}
Therefore, when $0<K<1$, $S^{\text{imp,}\chi^{\pm}}>S^{\text{imp,}M}>0$; when $1<K<2$, $S^{\text{imp,}\chi^{\pm}}<S^{\text{imp,}M}<0$; when $2<K<3$, $S^{\text{imp,}M}<S^{\text{imp,}C^{\pm}}<0$, $S^{\text{imp,}\chi^{\pm}}<S^{\text{imp,}C^{\pm}}<0$; when $K>3$, $S^{\text{imp,}M}<S^{\text{imp,}C^{\pm}}<S^{\text{imp,}\chi^{\pm}}<0$. These are all consistent with the RG flows obtained in Ref.~\onlinecite{PhysRevB.88.075131}, reproduced at the end of Appendix~\ref{sec:appRGflow} for convenience. We are also optimistic that Eq.~(\ref{RPAMimpent}) can be directly tested as an entanglement entropy against DMRG numerics,\cite{PhysRevLett.96.100603} possibly on the finite-size configuration proposed in Ref.~\onlinecite{PhysRevB.85.045120}.

\section{Conclusion and open questions\label{sec:openquestions}}

In this paper we have found the impurity entropy at a given RG fixed point of a junction of quantum wires by means of perturbation theory in the bulk interaction. Our results are respectively represented by Eqs.~(\ref{Oalphaimpent}) and (\ref{RPAimpent}) at the first order and in the RPA, in terms of the single-particle S-matrix at that fixed point. They are in full agreement with the $g$-theorem and the fermionic RG phase diagram.\cite{PhysRevB.84.155426,PhysRevB.88.075131} Our results mainly rest on two underlying assumptions: that at least some fixed points can be described by non-interacting single-particle S-matrices and a well-defined perturbation theory in interaction, and that no resonance exists near the Fermi level.

Although the second assumption is often easily realized (resonance typically requires fine-tuning model parameters), the implications of the first assumption is in fact not completely obvious. In particular, in very strongly attractive $Z_{3}$ symmetric Y-junctions, the fermionic RG approach based on the S-matrix of the original electrons and the RPA perturbation theory in interaction is clearly in conflict with results from bosonization. The $C^{\pm}$ fixed points are not predicted by bosonic methods; on the other hand, the $D$ fixed point involving Andreev scattering at the junction is postulated by bosonization,\cite{1742-5468-2006-02-P02008} but absent in the fermionic method which requires a particle-number conserving S-matrix. (Ref.~\onlinecite{PhysRevB.92.125138} obtains the $D$ fixed point through the re-fermionization of a bosonic Hamiltonian; the new free fermions there, however, are highly non-local with respect to the original electrons.) The possible existence of the $D$ fixed point hints at the breakdown of the perturbation theory at a more fundamental level for strongly attractive interactions.

The RPA itself deserves further comments. As shown by Eqs. (\ref{RPA2Dimpent}), (\ref{RPAA3impent}) and (\ref{RPAchiimpent}), at the geometrical fixed points accessible by both the RPA and the BCFT, the two theories find the same impurity entropy results. For the $D$ fixed point of a $Z_{2}$ symmetric 2-lead junction ($K_{1}=K_{2}$), we can show that this follows from the famous Dzyaloshinskii-Larkin theorem.\cite{SovPhysJETP.38.202,PhysRevB.44.12690} The theorem is originally formulated in the translation-invariant infinite-bandwidth Tomonaga-Luttinger model, whose left- and right-moving degrees of freedom have linear dispersion relations. It states that the sum of all Feynman diagrams which contain closed loops with more than two fermion lines, or equivalently closed loops connected to more than two forward-scattering interaction legs, should vanish after appropriate symmetrization. The theorem results from the strictly linear spectrum and the absence of backscattering; the latter leads to the conservation of chirality, i.e. left- and right-movers cannot turn into each other, and any fermion loop must have the same chirality for all its lines.

In the junction system the fermion chirality is clearly not conserved. In addition, without translational invariance, it becomes difficult to define the momentum (energy) carried by a forward-scattering interaction leg. The chirality problem may be formally circumvented as follows. Noticing that at the origin the right- and the left-movers are related by the S-matrix, $\psi_{jR}\left(  x=0\right) = \sum_{j^{\prime}} S_{jj^{\prime}}\psi_{j^{\prime}L}\left(  x=0\right)$, we adopt the ``unfolding'' trick in the non-interacting part of the system, and view the right-movers as linear combinations of left-movers analytically continued to the negative $x$-axis, so that the entire system contains left-movers only: $\psi_{jR}\left(  x\right) = \sum_{j^{\prime}} S_{jj^{\prime}}\psi_{j^{\prime}L}\left(  -x\right)$ ($x>0$). This is not how the unfolding trick is typically applied,\cite{giamarchi2003quantum} because the interaction Eq. (\ref{interaction0}) becomes seemingly non-local. However, the problem simplifies for the $N$ fixed point and the $Z_{2}$ symmetric $D$ fixed point. At the $N$ fixed point, $S_{jj^{\prime}} = \delta_{jj^{\prime}} e^{i \phi_{j}}$; the non-interacting part of the Hamiltonian is now diagonalized by the $\psi_{L}$ basis which is none other than the scattering basis, whereas the interaction can be represented by

\begin{equation}
H_{int}  =\sum_{j}g_{2}^{j}\int_{0}^{\infty}dx \psi^{\dag}_{jL}\left(  -x\right) \psi_{jL}\left(  -x\right) \psi^{\dag}_{jL}\left(  x\right) \psi_{jL}\left(  x\right) \text{.}%
\end{equation}
We can extend the domain of integration to $\left( -\infty, \infty \right) $ since the integrand is even. Going to the Fourier space then permits us to specify the momentum (energy) associated with an interaction leg:

\begin{equation}
H_{int}  =\frac{1}{2}\sum_{j}g_{2}^{j}\int\frac{dE_{1}dE_{2}dE_{3}dE_{4}}{2\pi
v_{Fj}}\delta\left(  E_{1}-E_{2}-E_{3}+E_{4}\right)  \psi_{jL}^{\dag}\left(  E_{1}\right)  \psi_{jL}\left(
E_{2}\right)  \psi_{jL}^{\dag}\left(  E_{3}\right)  \psi_{jL}\left(
E_{4}\right)\text{,}
\end{equation}
with the energy being $\left( E_{1}-E_{2} \right) $ in this case. One can finally apply the Dzyaloshinskii-Larkin theorem to each species of $\psi$. Meanwhile, at the $Z_{2}$ symmetric $D$ fixed point, the Dzyaloshinskii-Larkin theorem can be readily applied to the right- and left-movers of the infinite wire, which are themselves the scattering basis. We thus conclude that at these two fixed points the RPA becomes exact, i.e. the impurity entropy is given exclusively by diagrams with loops of only two fermion lines.

The argument above is heavily reliant on the real space integral evaluating to a delta function. Unfortunately, it fails not only at the non-geometrical fixed points where the right-mover densities generally contain cross-terms between different species of left-movers, but also at the $D$ fixed point of a $Z_{2}$ asymmetric 2-lead junction and the $\chi^{\pm}$ fixed points of a Y-junction where the parity symmetry is absent. The validity of the RPA therefore remains an important open question in general.

Whether the RPA fermion perturbation theory approach is to be justified or disproved away from the geometric fixed points, the natural next step is going beyond the RPA. The RG equations for the transmission probabilities have already been derived to the three-loop level;\cite{PhysRevB.80.045109,PhysRevB.84.155426} although the positions of the fixed points will generally change from the RPA result, it has been shown that in the $Z_{3}$ symmetric Y-junction neither the position of nor the scaling exponents at the $M$ fixed point are further corrected. We believe that an impurity entropy calculation beyond the RPA will shed more light on this issue when checked against the $g$-theorem.

\begin{acknowledgments}
This work was supported in part by NSERC of Canada, Discovery Grant 36318-2009. The author acknowledges inspirational discussions with Dr. Armin Rahmani, and is grateful to Prof. Ian Affleck for initiating the project and carefully proofreading the manuscript.
\end{acknowledgments}

\appendix

\section{Details of the RPA perturbation theory\label{sec:applnZ}}

In this Appendix, we derive Eq.~(\ref{RPAimpent}) from Eq.~(\ref{lnZm}).

After applying Wick's theorem and going to the frequency space, we find that two types of Matsubara sums are required for the ring diagrams. The first sum is standard\cite{mahan2000many} and occurs once for each fermion loop,

\begin{equation}
\frac{1}{\beta}\sum_{i\omega_{n}}\mathcal{G}_{j_{2}}\left(  E_{2},i\omega
_{n}\right)  \mathcal{G}_{j_{1}}\left(  E_{1},i\omega_{n}-ip_{m}\right)
=\frac{n_{F}\left(  E_{2}\right)  -n_{F}\left(  E_{1}\right)  }{-ip_{m}%
+E_{2}-E_{1}}\text{,} \label{bubbleMats}%
\end{equation}
where $p_{m}=2m\pi/\beta$ is a bosonic frequency. The second type of sums takes, for example, the following form

\begin{align}
&  \frac{1}{\beta}\sum_{ip_{m}}\frac{n_{F}\left(  E_{1}\right)  -n_{F}\left(
E_{2}\right)  }{ip_{m}+E_{1}-E_{2}}\frac{n_{F}\left(  E_{4}\right)
-n_{F}\left(  E_{3}\right)  }{ip_{m}+E_{4}-E_{3}}\frac{n_{F}\left(
E_{8}\right)  -n_{F}\left(  E_{7}\right)  }{ip_{m}+E_{8}-E_{7}}\nonumber\\
&  =\int\frac{d\tilde{\epsilon}}{2\pi i}\left[  n_{B}\left(  \tilde{\epsilon
}+i0\right)  \frac{n_{F}\left(  E_{1}\right)  -n_{F}\left(  E_{2}\right)
}{\tilde{\epsilon}+i0+E_{1}-E_{2}}\frac{n_{F}\left(  E_{4}\right)
-n_{F}\left(  E_{3}\right)  }{\tilde{\epsilon}+i0+E_{4}-E_{3}}\frac
{n_{F}\left(  E_{8}\right)  -n_{F}\left(  E_{7}\right)  }{\tilde{\epsilon
}+i0+E_{8}-E_{7}}\right. \nonumber\\
&  \left.  -n_{B}\left(  \tilde{\epsilon}-i0\right)  \frac{n_{F}\left(
E_{1}\right)  -n_{F}\left(  E_{2}\right)  }{\tilde{\epsilon}-i0+E_{1}-E_{2}%
}\frac{n_{F}\left(  E_{4}\right)  -n_{F}\left(  E_{3}\right)  }{\tilde
{\epsilon}-i0+E_{4}-E_{3}}\frac{n_{F}\left(  E_{8}\right)  -n_{F}\left(
E_{7}\right)  }{\tilde{\epsilon}-i0+E_{8}-E_{7}}\right]  \text{,}%
\end{align}
at the third order, where $n_{B}\left(  \epsilon\right)  =1/\left(  e^{\beta\epsilon}-1\right)  $ is the Bose distribution. This is proven by drawing a branch cut on the real axis, wrapping the contour of integration around the branch cut,\cite{mahan2000many} and calculating the contour integral

\begin{equation}
\oint \frac{dz}{2\pi i}n_{F}\left( z\right) \frac{1}{z+E_{1}-E_{2}}\frac{1}{%
z+E_{4}-E_{3}}\frac{1}{z+E_{8}-E_{7}}\text{.}
\end{equation}

To proceed further we must integrate over the loop energies, e.g. $E_{1}$ and $E_{2}$. This is made possible by extending the limits of integration to $\pm \infty$, employing residue techniques,\cite{SovPhysJETP.38.202} and keeping track of the phase factors due to the interaction vertices. The terms proportional to $n_{F}\left(  E_{1}\right)$ and $n_{F}\left(  E_{2}\right)$ should have $E_{2}$ and $E_{1}$ integrated over, respectively; the two terms are then combined using the relation $\tilde{\epsilon}\pm i0+E_{1}-E_{2}=0$ as appropriate. For each fermion loop these procedures yield one factor of $\tilde{\epsilon}$.

Once the loop energy integrals are done, we need to complete the real space integrals which accompany the interaction vertices. Our tools are the following equations:

\begin{equation}
\int_{0}^{\infty}d\tilde{y}_{1}e^{2iE^{+}\tilde{y}_{1}}=\frac{i}{2E^{+}}\text{,}
\label{ordrsint}%
\end{equation}
and

\begin{equation}
\prod_{l=1}^{2M+1}\left(  \int_{0}^{\infty
}d\tilde{y}_{l}\right)  e^{iE^{+}\left(  \tilde{y}_{1}+\tilde{y}%
_{2M+1}\right)  }\prod_{j=1}^{M}\left[  e^{iE^{+}\left(  \tilde{y}%
_{2j-1}+\tilde{y}_{2j+1}-2\tilde{y}_{2j}\right)  }\theta\left(  \tilde
{y}_{2j-1}-\tilde{y}_{2j}\right)  \theta\left(  \tilde{y}_{2j+1}-\tilde
{y}_{2j}\right)  \right]  =\left(  \frac{i}{2E^{+}}\right)  ^{2M+1}C_{M}\text{,}
\label{catalanintegral}%
\end{equation}
where $E^{+}\equiv E+i0$, and $C_{M}=\left(  2M\right)  !/\left[  M!\left(  M+1\right)  !\right]  $ is the $M$th Catalan number;\cite{OEIS.A000108,*OEIS.A008315,2016arXiv160100510S} the first six Catalan numbers $0\leq M\leq5$ are $1$, $1$, $2$, $5$, $14$, $42$. The Heaviside $\theta$ functions in Eq.~(\ref{catalanintegral}) originate from the loop energy integrals.

At the $m$th order in interaction ($m>1$), the $m$ factors of $1/\tilde{\epsilon}$ from real space integrals are paired with the $m$ factors of $\tilde{\epsilon}$ from loop energy integrals. The $\tilde{\epsilon}$ integral is then straightforward,

\begin{align}
&  \int_{-D}^{D}d\tilde{\epsilon}\,n_{B}\left(  \tilde{\epsilon}+i0\right)
\nonumber\\
&  =\frac{1}{\beta}\left(  \ln\left[  1-e^{-\beta\left(  D+i0\right)
}\right]  -\ln\left[  1-e^{-\beta\left(  -D+i0\right)  }\right]  \right)
=-\frac{i\pi}{\beta}-D\text{.}%
\end{align}
As before, the constant term $-D$ shifts the ground state energy and does not
contribute to the impurity entropy. Collecting terms of all orders we can
write the impurity part of $\delta\ln Z$ in the RPA as%

\begin{align}
\delta\ln Z\left(  \lambda\right)   &  \rightarrow-\frac{1}{2}\left\{
\lambda\sum_{n_{1}}\frac{\alpha_{n_{1}}}{2}W_{n_{1}n_{1}}+\frac{\lambda^{2}%
}{2}\sum_{n_{1}n_{2}}\frac{\alpha_{n_{1}}}{2}\frac{\alpha_{n_{2}}}{2}%
W_{n_{1}n_{2}}W_{n_{2}n_{1}}\right. \nonumber\\
&  +\frac{\lambda^{3}}{3}\sum_{n_{1}n_{2}n_{3}}\frac{\alpha_{n_{1}}}{2}%
\frac{\alpha_{n_{2}}}{2}\frac{\alpha_{n_{3}}}{2}\left[  \delta_{n_{1}n_{2}%
}\delta_{n_{2}n_{3}}W_{n_{3}n_{1}}\right. \nonumber\\
&  \left.  +W_{n_{1}n_{2}}\delta_{n_{2}n_{3}}\delta_{n_{3}n_{1}}+\delta
_{n_{1}n_{2}}W_{n_{2}n_{3}}\delta_{n_{3}n_{1}}+W_{n_{1}n_{2}}W_{n_{2}n_{3}%
}W_{n_{3}n_{1}}\right]  +\cdots\Bigg\}\text{.} \label{RPAthedynpot}%
\end{align}
The auxiliary variable $\lambda$ is introduced for power-counting of $\alpha$ and is set to unity in the end. The rules to enumerate terms in Eq.~(\ref{RPAthedynpot}) are as follows. At $O\left(  \alpha^{m}\right)  $, there is a total of $m$ factors of $\delta$ and $W$ in every term; the $\delta$ factors always appear in even-length strings separated by the $W$ factors. For each string of $\delta$ factors of length $2M$ (where $M$ is an integer) that it contains, a term is multiplied by a Catalan number prefactor $C_{M}$. Most importantly, all terms in Eq.~(\ref{RPAthedynpot}) are interpreted to be cyclic: at $O\left(  \alpha^{5}\right)  $, for instance, $\delta W\delta\delta\delta$ and $\delta\delta W\delta\delta$ are both allowed and count as different terms, and both are understood to contain one string of $\delta\delta\delta\delta$ with the prefactor of $C_{4}=14$.

To resum the series in Eq.~(\ref{RPAthedynpot}) it is convenient to take the $\lambda$ derivative. Taking the cyclicity into account, every term in Eq.~(\ref{RPAthedynpot}) may be represented as the cyclic concatenation of a (possibly empty) even-length string of $\delta$, and another non-cyclic string which starts with a $W$, ends with a $W$, and contains only even-length strings of $\delta$ factors with multiplicative Catalan number prefactors; e.g. the term $\delta WW\delta\delta W\delta\delta\delta$ is seen as $\delta\delta\delta\delta$ attached to $WW\delta\delta W$. Equivalently,

\begin{equation}
\frac{\partial\delta\ln Z\left(  \lambda\right)  }{\partial\lambda}=-\frac
{1}{2}\operatorname*{tr}\left[  \frac{1}{\lambda}\frac{\bar{\Pi}^{\prime
}\left(  \lambda\right)  }{2}\left(  W+W\frac{\Pi\left(  \lambda\right)  }%
{2}W\right)  \right]  \text{.}%
\end{equation}
Here $\mathcal{Q}$ has been defined in Eq.~(\ref{scriptQ}), $\Pi\left(  \lambda\right)  $ is found in Ref.~\onlinecite{2016arXiv160100510S} and given by

\begin{equation}
\Pi  \equiv2\left(  \mathcal{Q}  -W\right)  ^{-1}\text{,} \label{RPAvertex}%
\end{equation}
with $\alpha$ replaced by $\lambda\alpha$ in $\mathcal{Q}$, and

\begin{align}
\frac{\bar{\Pi}_{jj^{\prime}}^{\prime}\left(  \lambda\right)  }{2}  &
\equiv\frac{\lambda\alpha_{j}}{2}\sum_{M=0}^{\infty}\left(  2M+1\right)
C_{M}\left(  \frac{\lambda\alpha_{j}}{2}\right)  ^{2M}\delta_{jj^{\prime}%
}\nonumber\\
&  =\lambda\frac{\partial}{\partial\lambda}\left(  \frac{\lambda\alpha_{j}%
}{1+\sqrt{1-\lambda^{2}\alpha_{j}^{2}}}\right)  \delta_{jj^{\prime}}\text{;}%
\end{align}
the factor of $2M+1$ is the number of ways to partition a string of length $2M$ into two, because e.g. for $M=1$, $\delta\delta W\cdots W$, $\delta W\cdots W\delta$ and $W\cdots W\delta\delta$ all contribute equally to the
series. Performing the $\lambda$ integration then leads to

\begin{equation}
\delta\ln Z\left(  \lambda\right)  \rightarrow\frac{1}{2}\ln\det\left(
\delta_{n_{1}n_{2}}-\frac{\lambda\alpha_{n_{1}}}{1+\sqrt{1-\lambda^{2}%
\alpha_{n_{1}}^{2}}}W_{n_{1}n_{2}}\right)  \text{.}%
\end{equation}
Taking $\lambda=1$ and subtracting with the $N$ fixed point value, we eventually recover Eq.~(\ref{RPAimpent}).

\section{S-matrix RG equation and fixed points for 2-lead junctions and Y-junctions\label{sec:appRGflow}}

In this Appendix we explicitly write down the S-matrix RG equations specific to 2-lead junctions and Y-junctions, both at the first order\cite{PhysRevLett.71.3351,PhysRevB.66.165327} and in the
RPA.\cite{0295-5075-82-2-27001,PhysRevB.84.155426,PhysRevB.88.075131,2016arXiv160100510S} We show that in all these cases it is possible to eliminate the phases of transmission/reflection amplitudes, resulting in a set of equations containing only the transmission/reflection probability matrix $W$. The fixed points of these equations are then listed and their local stability analyzed, for comparison with the impurity entropy results in Section~\ref{sec:gtheorem}.

\subsection{First order in interaction\label{subsec:appRGflowOalpha}}

To the first order in interaction, for a generic junction of $N$ quantum wires which is not on resonance, the S-matrix $S\left(  D\right)  $ obeys the following RG equation as the running cutoff $D$ is reduced:\cite{PhysRevB.66.165327}

\begin{equation}
-\frac{dS_{jj^{\prime}}\left(  D\right)  }{d\ln D}=-\frac{1}{2}\sum_{n=1}^{N}%
\alpha_{n}\left[  S_{jn}\left(  D\right)  S_{nn}^{\ast
}\left(  D\right)  S_{nj^{\prime}}\left(  D\right)  -\delta_{j^{\prime}%
n}S_{nn}\left(  D\right)  \delta_{nj}\right]  \text{.}
\label{NazarovGlazman}%
\end{equation}
This can be derived by, for example, considering the two-point correlation function between left- and right-movers on different wires.\cite{PhysRevB.68.035421} To lighten notations we suppress the $D$ dependence in the following.

\subsubsection{2-lead junction}

For the transmission amplitude $S_{12}$, Eq.~(\ref{NazarovGlazman}) becomes

\begin{equation}
-\frac{dS_{12}}{d\ln D}=-\frac{1}{2}\left(  \alpha_{1}W_{11}+\alpha_{2}%
W_{22}\right)  S_{12}\text{.}%
\end{equation}
In the 2-lead junction, unitarity implies $W_{11}=W_{22}=1-W_{12}$. Therefore, we have the following equation for $W_{12}\equiv\left\vert S_{12}\right\vert  ^{2}$,

\begin{equation}
-\frac{dW_{12}}{d\ln D}=-\left(  \alpha_{1}+\alpha_{2}\right)  W_{12}\left(
1-W_{12}\right)  \text{.} \label{Oalpha2leadprob}%
\end{equation}

In the vicinity of the complete reflection fixed point $N$ ($W_{12}=0$), linearizing Eq.~(\ref{Oalpha2leadprob}), we find $-dW_{12}/d\ln D\approx -\left(  \alpha_{1}+\alpha_{2}\right)  W_{12}$; thus the $N$ fixed point has a scaling exponent for the conductance $-\left(  \alpha_{1}+\alpha_{2}\right) $, and is stable if $-\left(  \alpha_{1}+\alpha_{2}\right)  <0$ and unstable if $-\left(  \alpha_{1}+\alpha_{2}\right)  >0$. Similarly, the perfect transmission fixed point $D$ ($W_{12}=1$) has a conductance scaling exponent $\alpha_{1}+\alpha_{2}$, and is stable if $\alpha_{1}+\alpha_{2}<0$ and unstable if $\alpha_{1}+\alpha_{2}>0$.

\subsubsection{Y-junction}

For a Y-junction, from Eq.~(\ref{NazarovGlazman})

\begin{equation}
-\frac{dS_{12}}{d\ln D}=-\frac{1}{2}\left(  \alpha_{1}W_{11}S_{12}+\alpha
_{2}W_{22}S_{12}+\alpha_{3}S_{13}S_{33}^{\ast}S_{32}\right)  \text{.}%
\end{equation}
To reduce this to an equation of $W$ only, we need to relate the product $S_{12}^{\ast}S_{13}S_{33}^{\ast}S_{32}$ to $W$. This is achieved by taking advantage of unitarity of the S-matrix:%

\begin{equation}
S_{12}^{\ast}S_{13}S_{33}^{\ast}S_{32}+\text{c.c.}=W_{23}W_{22}-W_{13}%
W_{12}-W_{33}W_{32}\text{.}%
\end{equation}
Thus, the RG equation obeyed by $W_{12}$ takes the form

\begin{equation}
-\frac{dW_{12}}{d\ln D}=-\left(  \alpha_{1}W_{11}+\alpha_{2}W_{22}\right)
W_{12}-\frac{1}{2}\alpha_{3}\left(  W_{23}W_{22}-W_{13}W_{12}-W_{33}%
W_{32}\right)  \text{.}%
\end{equation}
Unitarity dictates that there are only $4$ independent matrix elements of $W$. Following Refs.~\onlinecite{PhysRevB.84.155426,PhysRevB.88.075131}, we parametrize the $W$ matrix by four real numbers $\left(  a,b,c,\bar{c}\right)  $ as follows:

\begin{equation}
W=\frac{1}{6}\left(
\begin{array}
[c]{ccc}%
2+3a+b-\sqrt{3}\left(  c+\bar{c}\right)  & 2-3a+b-\sqrt{3}\left(  c-\bar
{c}\right)  & 2\left(  1-b+\sqrt{3}c\right) \\
2-3a+b+\sqrt{3}\left(  c-\bar{c}\right)  & 2+3a+b+\sqrt{3}\left(  c+\bar
{c}\right)  & 2\left(  1-b-\sqrt{3}c\right) \\
2\left(  1-b+\sqrt{3}\bar{c}\right)  & 2\left(  1-b-\sqrt{3}\bar{c}\right)  &
2\left(  1+2b\right)
\end{array}
\right)  \text{.} \label{abccbar}%
\end{equation}
In the presence of time-reversal symmetry, $c=\bar{c}$. If wires $1$ and $2$ are symmetrically coupled to the junction, $c=-\bar{c}$; we further find $a=b$ if $Z_{3}$ symmetry exists for the non-interacting system.

Eq.~(\ref{NazarovGlazman}) and Eq.~(\ref{abccbar}) now lead to a closed set of equations for $a$, $b$, $c$ and $\bar{c}$,%

\begin{subequations}
\label{Oalpha3leadprob}%
\begin{align}
-\frac{da}{d\ln D}  &  =\frac{1}{12}\left[  \left(  \alpha_{1}+\alpha
_{2}\right)  \left(  3+a+3b-6a^{2}-ab-c\bar{c}\right)  -\left(  \alpha
_{1}-\alpha_{2}\right)  \sqrt{3}\left(  1-2a\right)  \left(  c+\bar{c}\right)
\right] \nonumber\\
&  +\frac{1}{3}\alpha_{3}\left(  a-ab-c\bar{c}\right)  \text{,}%
\end{align}

\begin{align}
-\frac{db}{d\ln D}  &  =\frac{1}{12}\left[  \left(
\alpha_{1}+\alpha_{2}\right)  \left(  1+3a+b-2b^{2}-3ab-3c\bar{c}\right)
+\left(  \alpha_{1}-\alpha_{2}\right)  \sqrt{3}\left(  1+2b\right)  \left(
c+\bar{c}\right)  \right] \nonumber\\
&  +\frac{1}{3}\alpha_{3}\left(  1+b-2b^{2}\right)  \text{,}%
\end{align}

\begin{align}
-\frac{dc}{d\ln D}  &  =\frac{1}{12}\left[  -\left(
\alpha_{1}+\alpha_{2}\right)  \left(  c\left(  1+2b+6a\right)  -3\bar
{c}\right)  +\sqrt{3}\left(  \alpha_{1}-\alpha_{2}\right)  \left(
-1-a+b+ab+2c^{2}+c\bar{c}\right)  \right] \nonumber\\
&  -\frac{1}{3}\alpha_{3}\left(  1+2b\right)  c\text{,}%
\end{align}

\begin{align}
-\frac{d\bar{c}}{d\ln D}  &  =\frac{1}{12}\left[
-\left(  \alpha_{1}+\alpha_{2}\right)  \left(  \bar{c}\left(  1+2b+6a\right)
-3c\right)  +\sqrt{3}\left(  \alpha_{1}-\alpha_{2}\right)  \left(
-1-a+b+ab+2\bar{c}^{2}+c\bar{c}\right)  \right] \nonumber\\
&  -\frac{1}{3}\alpha_{3}\left(  1+2b\right)  \bar{c}\text{.}%
\end{align}
\end{subequations}

After finding a fixed point $\left(  a_{0},b_{0},c_{0},\bar{c}_{0}\right)  $ of Eq.~(\ref{Oalpha3leadprob}), we can again linearize the equations\cite{PhysRevB.84.155426,PhysRevB.88.075131} by expanding in terms of small deviations from the fixed point, $\mathbf{x}\equiv\left(  a-a_{0},b-b_{0},c-c_{0},\bar{c}-\bar{c}_{0}\right)  $:

\begin{equation}
-\frac{d\mathbf{x}}{d\ln D}=\mathbf{Mx}\text{.} \label{linearization}%
\end{equation}
The $4\times4$ matrix $\mathbf{M}$ have eigenvalues $\lambda_{l}$ with corresponding left eigenvectors $\mathbf{v}_{l}$, $l=1$, $2$, $3$, $4$. For an RG flow starting in the vicinity of the fixed point in question, the solution to Eq.~(\ref{linearization}) takes the form

\begin{equation}
\mathbf{x}\left(  D\right)  =\sum_{j=1}^{4}C_{l}\left(  \frac{D_{0}}%
{D}\right)  ^{\lambda_{l}}\mathbf{v}_{l}\text{,} \label{linsolution}%
\end{equation}
where $C_{j}$ are constants and $D_{0}$ is the ultraviolet cutoff. $\lambda_{l}$ thus controls the stability of the fixed point: $\mathbf{v}_{l}$ is a stable scaling direction if $\lambda_{l}<0$, and an unstable one if $\lambda_{l}>0$. For a junction attached to FL leads, replacing $D$ by the temperature $T$ in Eq.~(\ref{linsolution}), we will find the low temperature conductance at a stable fixed point with all $\lambda_{l}<0$, or the high temperature conductance at a completely unstable fixed point with all $\lambda_{l}>0$; thus $\lambda_{l}$ are the scaling exponents of the conductance. ($\lambda_{l}$ are generally not the same with the S-matrix scaling exponents discussed in Ref.~\onlinecite{PhysRevB.66.165327}.)

In the following we list $\lambda_{l}$ for the first order fixed points and discuss their physical meanings.

$N$\textbf{ fixed point}: $\left(  a_{0},b_{0},c_{0},\bar{c}_{0}\right)  =\left(  1,1,0,0\right)  $; $\lambda_{N\text{,}1}=-\left(  \alpha_{2}+\alpha_{3}\right)  $, $\lambda_{N\text{,}2}=-\left(  \alpha_{3}+\alpha_{1}\right)  $, $\lambda_{N\text{,}3}=-\left(  \alpha_{1}+\alpha_{2}\right)  $ and $\lambda_{N\text{,}4}=-\left(  \alpha_{1}+\alpha_{2}+\alpha_{3}\right)  $, with eigenvectors $\mathbf{v}_{N\text{,}1}=\left(  1,-1/3,0,0\right)  $, $\mathbf{v}_{N\text{,}2}=\left(  0,-2/\sqrt{3},1,1\right)  $, $\mathbf{v}_{N\text{,}3}=\left(  0,2/\sqrt{3},1,1\right)  $ and $\mathbf{v}_{N\text{,}4}=\left(  0,0,1,-1\right)  $ respectively. $\lambda_{N\text{,}1}$ corresponds to the process where a single electron tunnels between wires $2$ and $3$; thus we know from the 2-lead junction problem that it controls the RG flow between $N$ and $A_{1}$. (By a flow ``between'' two fixed points, we refer to a flow which, starting sufficiently close to either of the two fixed points, can come into arbitrary proximity to the other.) Similarly, depending on the attractive or repulsive nature of the interactions, $\lambda_{N\text{,}2}$ and $\lambda_{N\text{,}3}$ control the RG flow from/to $A_{2}$ and $A_{3}$, respectively. The flow between $N$ and $M$ is jointly controlled by $\lambda_{N\text{,}1}$, $\lambda_{N\text{,}2}$ and $\lambda_{N\text{,}3}$.
On the other hand, along the direction of $\mathbf{v}_{N\text{,}4}$, $c=-\bar{c}\neq0$; thus $\mathbf{v}_{N\text{,}4}$ represents a chiral perturbation, and $\lambda_{N\text{,}4}$ controls the flows from/to $\chi_{\pm}$ which are the only fixed points breaking time-reversal symmetry at the first order.

We note that $a$, $b$, $c$ and $\bar{c}$ are subject to additional constraints imposed by the S-matrix unitarity.\cite{PhysRevB.84.155426,PhysRevB.88.075131} By considering physically allowed S-matrices, it can be shown that
$C_{N\text{,}4}=0$ in Eq.~(\ref{linsolution}) (i.e. $\mathbf{v}_{N\text{,}4}$ is not allowed) unless $C_{N\text{,}1}$, $C_{N\text{,}2}$ and $C_{N\text{,}3}$ are all nonzero. For this reason, $\lambda_{N\text{,}4}$ has not been regarded as an independent conductance scaling exponent in Refs.~\onlinecite{PhysRevB.84.155426,PhysRevB.88.075131}. Intuitively, this can also be understood from the fact that the breaking of time-reversal symmetry ($\mathbf{v}_{N\text{,}4}$) requires the presence of single electron tunneling between all three wires ($\mathbf{v}_{N\text{,}1}$, $\mathbf{v}_{N\text{,}2}$ and $\mathbf{v}_{N\text{,}3}$), so that a magnetic flux threaded into the junction cannot be trivially gauged away. It should be also mentioned that $\lambda_{N\text{,}4}$ is never the leading scaling exponent in either the high temperature or the low temperature limit. For instance, if we assume $\lambda_{N\text{,}4}$ is the leading exponent at low temperatures, then $\lambda_{N\text{,}4}$ must be greater than all remaining $\lambda$'s, and we find all $\alpha$'s are negative and $N$ is unstable in all directions, which contradicts our assumption.

$A_{j}$\textbf{ fixed points}: At $A_{1,2}$, $\left(  a_{0},b_{0},c_{0},\bar{c}_{0}\right)  =\left(  1/2,-1/2,\mp\sqrt{3}/2,\mp\sqrt{3}/2\right)  $; at $A_{3}$, $\left(  a_{0},b_{0},c_{0},\bar{c}_{0}\right)  =\left(-1,1,0,0\right)  $. We now focus on $A_{3}$ where wire $3$ is decoupled, and wires $1$ and $2$ are perfectly connected.

At $A_{3}$, $\lambda_{A_{3}\text{,}1}=\lambda_{A_{3}\text{,}2}=-\alpha_{3}$, $\lambda_{A_{3}\text{,}3}=\alpha_{1}+\alpha_{2}$ and $\lambda_{A_{3}\text{,}4}=\left(  \alpha_{1}+\alpha_{2}-2\alpha_{3}\right)  /2$, with eigenvectors

\begin{subequations}
\begin{equation}
\mathbf{v}_{A_{3}\text{,}1}=\left(  0,0,1,-1\right)  \text{,}%
\end{equation}

\begin{equation}
\mathbf{v}_{A_{3}\text{,}2}=\left(  0,-\left(  2/\sqrt{3}\right)  \left(
\alpha_{1}+\alpha_{2}\right)  ,\alpha_{1}-\alpha_{2},\alpha_{1}-\alpha
_{2}\right)  \text{,}%
\end{equation}

\begin{equation}
\mathbf{v}_{A_{3}\text{,}3}=\left(  -\sqrt{3}\left(  \alpha_{1}+\alpha
_{2}+2\alpha_{3}\right)  ,-\left(  \alpha_{1}+\alpha_{2}+2\alpha_{3}\right)
/\sqrt{3},\alpha_{1}-\alpha_{2},\alpha_{1}-\alpha_{2}\right)  \text{,}%
\end{equation}

\begin{equation}
\mathbf{v}_{A_{3}\text{,}4}=\left(  0,0,1,1\right)
\end{equation}
\end{subequations}
respectively. Again $\lambda_{A_{3}\text{,}1}$ corresponds to the single-electron tunneling between wires $2$ and $3$, $\lambda_{A_{3}\text{,}2}$ to that between $3$ and $1$, and $\lambda_{A_{3}\text{,}3}$ to that between $1$ and $2$. As before $\lambda_{A_{3}\text{,}3}$ controls the flow from/to $N$, and we see from the eigenvector $\mathbf{v}_{A_{3}\text{,}1}$ that $\lambda_{A_{3}\text{,}1}=\lambda_{A_{3}\text{,}2}$ controls the flows from/to $\chi_{\pm}$.

In the special case of 1-2 symmetric interaction, $\alpha_{1}=\alpha_{2}$, it is clear that $\mathbf{v}_{A_{3}\text{,}4}$ is the only scaling direction breaking the 1-2 symmetry. Now the flows from/to $A_{1}$ and $A_{2}$ are controlled by $\lambda_{A_{3}\text{,}4}$ alone, and the flow from/to $M$ is controlled by $\lambda_{A_{3}\text{,}1}=\lambda_{A_{3}\text{,}2}$ and $\lambda_{A_{3}\text{,}3}$ together but not $\lambda_{A_{3}\text{,}4}$.

We observe that $\mathbf{v}_{A_{3}\text{,}1}$ and $\mathbf{v}_{A_{3}\text{,}4}$ break the time-reversal symmetry and the 1-2 symmetry of the junction respectively without changing $a$ or $b$. This is again forbidden by unitarity\cite{PhysRevB.84.155426,PhysRevB.88.075131} at $A_{3}$, i.e. $C_{A_{3}\text{,}1}=C_{A_{3}\text{,}4}=0$ in Eq.~(\ref{linsolution}) unless either $C_{A_{3}\text{,}2}$ or $C_{A_{3}\text{,}3}$ is nonzero. Physically it reflects the fact that any time-reversal asymmetry or 1-2 asymmetry at the junction should introduce perturbations that interrupt the perfectly connected wire. $\lambda_{A_{3}\text{,}4}$ is therefore not treated as a scaling exponent in Refs.~\onlinecite{PhysRevB.84.155426,PhysRevB.88.075131}.

$\chi^{\pm}$\textbf{ fixed points}: At $\chi^{\pm}$, $\left(  a_{0},b_{0},c_{0},\bar{c}_{0}\right)  =\left(  -1/2,-1/2,\pm\sqrt{3}/2,\mp\sqrt{3}/2\right)  $; $\lambda_{\chi^{\pm},1}=\alpha_{1}$, $\lambda_{\chi^{\pm},2}=\alpha_{2}$, $\lambda_{\chi^{\pm},3}=\alpha_{3}$ and $\lambda_{\chi^{\pm}\text{,}4}=\left(  \alpha_{1}+\alpha_{2}+\alpha_{3}\right)  /2$, with eigenvectors

\begin{subequations}
\begin{align}
\mathbf{v}_{\chi^{\pm}\text{,}1}  &  =\left(  -\sqrt{3}\alpha_{1},-\left(
\alpha_{1}+2\left(  \alpha_{2}+\alpha_{3}\right)  \right)  /\sqrt{3},\left(
\alpha_{1}-\left(  \alpha_{2}+\alpha_{3}\right)  \mp\left(  \alpha_{2}%
+\alpha_{3}\right)  \right)  ,\right. \nonumber\\
&  \left.  \left(  \alpha_{1}-\left(  \alpha_{2}+\alpha_{3}\right)  \pm\left(
\alpha_{2}+\alpha_{3}\right)  \right)  \right)  \text{,}%
\end{align}

\begin{align}
\mathbf{v}_{\chi^{\pm}\text{,}2}  &  =\left(  \sqrt{3}\alpha_{2},\left(
\alpha_{2}+2\left(  \alpha_{1}+\alpha_{3}\right)  \right)  /\sqrt{3},\left(
\alpha_{2}-\left(  \alpha_{1}+\alpha_{3}\right)  \pm\left(  \alpha_{1}%
+\alpha_{3}\right)  \right)  ,\right. \nonumber\\
&  \left.  \left(  \alpha_{2}-\left(  \alpha_{1}+\alpha_{3}\right)  \mp\left(
\alpha_{1}+\alpha_{3}\right)  \right)  \right)  \text{,}%
\end{align}

\begin{equation}
\mathbf{v}_{\chi^{\pm}\text{,}3}=\left(  \sqrt{3}\left(  \alpha_{1}+\alpha
_{2}\right)  ,-\left(  \alpha_{1}+\alpha_{2}-4\alpha_{3}\right)  /\sqrt{3}%
,\pm\left(  \alpha_{1}+\alpha_{2}\right)  ,\mp\left(  \alpha_{1}+\alpha
_{2}\right)  \right)  \text{,}%
\end{equation}

\begin{equation}
\mathbf{v}_{\chi^{\pm}\text{,}4}=\left(  \sqrt{3},\sqrt{3},\pm1,\mp1\right)
\end{equation}
\end{subequations}
respectively. $\lambda_{\chi^{\pm}\text{,}1}$ corresponds to the single-electron tunneling between wires $2$ and $3$, and controls the flows from/to $A_{1}$; similarly $\lambda_{\chi^{\pm}\text{,}2}$ and $\lambda_{\chi^{\pm}\text{,}3}$ controls the flows from/to $A_{2}$ and $A_{3}$ respectively. $\lambda_{\chi^{\pm}\text{,}4}$ controls the RG flows from/to $N$ and $M$.

The scaling direction $\mathbf{v}_{\chi^{\pm}\text{,}4}$\ is forbidden by unitarity at $\chi^{\pm}$ ($C_{\chi^{\pm}\text{,}4}=0$ in Eq.~(\ref{linsolution}) unless $C_{\chi^{\pm}\text{,}1}$, $C_{\chi^{\pm}\text{,}2}$ and $C_{\chi^{\pm}\text{,}3}$ are all nonzero), so once more $\lambda_{\chi^{\pm}\text{,}4}$ is not treated as a scaling exponent in Refs.~\onlinecite{PhysRevB.84.155426,PhysRevB.88.075131}. In terms of the mapping to the dissipative Hofstadter model,\cite{1742-5468-2006-02-P02008} $\chi^{\pm}$ correspond to the localized phases of a quantum Brownian particle subject to a magnetic field and a triangular-lattice potential; $\lambda_{\chi^{\pm}\text{,}1}$, $\lambda_{\chi^{\pm}\text{,}2}$ and $\lambda_{\chi^{\pm}\text{,}3}$ are then due to instantons tunneling back and forth between the three inequivalent nearest neighbor pairs of potential minima, while $\lambda_{\chi^{\pm}\text{,}4}$ arises from instantons tunneling along the edges of elementary triangles formed by the potential minima. It is therefore reasonable that $\mathbf{v}_{\chi^{\pm}\text{,}4}$ is allowed only if there exist deviations from $\chi^{\pm}$ along all three remaining scaling directions $\mathbf{v}_{\chi^{\pm}\text{,}1}$, $\mathbf{v}_{\chi^{\pm}\text{,}2}$ and $\mathbf{v}_{\chi^{\pm}\text{,}3}$.

$M$\textbf{ fixed point}: Due to time-reversal symmetry $c_{0}=\bar{c}_{0}=0$; it is straightforward to find $a_{0}$ and $b_{0}$ from Eqs.~(\ref{OalphaMW}) and (\ref{abccbar}). There are $4$ scaling exponents at $M$:%

\begin{subequations}
\label{OalphaMexps}
\begin{equation}
\lambda_{M\text{,}1}=-\frac{\alpha_{1}\alpha_{2}\alpha_{3}}{\alpha_{1}%
\alpha_{2}+\alpha_{2}\alpha_{3}+\alpha_{3}\alpha_{1}}\text{,}%
\end{equation}

\begin{equation}
\lambda_{M\text{,}2}=\frac{\alpha_{1}^{2}\left(  \alpha
_{2}+\alpha_{3}\right)  +\alpha_{2}^{2}\left(  \alpha_{3}+\alpha_{1}\right)
+\alpha_{3}^{2}\left(  \alpha_{1}+\alpha_{2}\right)  }{\alpha_{1}\alpha
_{2}+\alpha_{2}\alpha_{3}+\alpha_{3}\alpha_{1}}\text{,}%
\end{equation}

\begin{align}
\lambda_{M\text{,}3\left(  4\right)  }  &  =\frac{1}%
{4}\left(  \alpha_{1}\alpha_{2}+\alpha_{2}\alpha_{3}+\alpha_{3}\alpha
_{1}\right)  ^{-1}\left\{  \left(  \alpha_{1}+\alpha_{2}\right)  \left(
\alpha_{2}+\alpha_{3}\right)  \left(  \alpha_{3}+\alpha_{1}\right)  \right.
\nonumber\\
&  \left.  \pm\sqrt{\left(  \alpha_{1}+\alpha_{2}\right)  ^{2}\left(
\alpha_{2}+\alpha_{3}\right)  ^{2}\left(  \alpha_{3}+\alpha_{1}\right)
^{2}-8\alpha_{1}\alpha_{2}\alpha_{3}\left(  \alpha_{1}+\alpha_{2}\right)
\left(  \alpha_{2}+\alpha_{3}\right)  \left(  \alpha_{3}+\alpha_{1}\right)
}\right\}  \text{.}%
\end{align}
\end{subequations}
Whenever $M$ exists [see the conditions below Eq.~(\ref{OalphaMW})], $\lambda_{M\text{,}3}$, $\lambda_{M\text{,}4}$ are both real. Note that, unlike the situations at $N$, $A_{j}$ and $\chi^{\pm}$, at $M$ the four scaling directions are fully independent of each other. In the special case of $Z_{3}$ symmetric interactions ($\alpha_{j}=\alpha$), $\lambda_{M\text{,}1}=-\alpha/3$, $\lambda_{M\text{,}2}=2\alpha$, $\lambda_{M\text{,}3}=\lambda_{M\text{,}4}=2\alpha/3$, in agreement with the conductance predictions of Ref.~\onlinecite{PhysRevB.66.165327}. The corresponding left eigenvectors are

\begin{subequations}
\begin{equation}
\mathbf{v}_{M\text{,}1}=\left(  0,0,1,-1\right)  \text{,}%
\end{equation}

\begin{equation}
\mathbf{v}_{M\text{,}2}=\left(  -\sqrt{3}\left(  \alpha_{1}+\alpha_{2}\right)
,-\left(  \alpha_{1}+\alpha_{2}+4\alpha_{3}\right)  /\sqrt{3},\alpha
_{1}-\alpha_{2},\alpha_{1}-\alpha_{2}\right)  \text{.}%
\end{equation}
\end{subequations}

We can infer from the form of $\mathbf{v}_{M\text{,}1}$ that $\lambda_{M\text{,}1}$ controls the flows from/to $\chi^{\pm}$. $\mathbf{v}_{M\text{,}3}$ and $\mathbf{v}_{M\text{,}4}$ are too complicated to be given here in general.

In the case of 1-2 symmetric interactions ($\alpha_{1}=\alpha_{2}$), significant simplifications occur:%

\begin{subequations}
\begin{equation}
\lambda_{M\text{,}1}=-\frac{\alpha_{1}\alpha_{3}}{\alpha_{1}+2\alpha_{3}%
}\text{, }\lambda_{M\text{,}2}=\frac{2\left(  \alpha_{1}^{2}+\alpha_{1}%
\alpha_{3}+\alpha_{3}^{2}\right)  }{\alpha_{1}+2\alpha_{3}}\text{;}%
\end{equation}

\begin{equation}
\lambda_{M\text{,}3}=\frac{\alpha_{3}\left(  \alpha_{1}+\alpha_{3}\right)
}{\alpha_{1}+2\alpha_{3}}\text{, }\lambda_{M\text{,}4}=\frac{\alpha_{1}\left(
\alpha_{1}+\alpha_{3}\right)  }{\alpha_{1}+2\alpha_{3}}\text{;}%
\end{equation}

\end{subequations}
\begin{subequations}
\begin{equation}
\mathbf{v}_{M\text{,}3}=\left(  3\alpha_{3}\left(  \alpha_{1}+\alpha
_{3}\right)  ,-2\alpha_{1}\left(  \alpha_{1}+2\alpha_{3}\right)  ,0,0\right)
\text{,}%
\end{equation}

\begin{equation}
\mathbf{v}_{M\text{,}4}=\left(  0,0,1,1\right)  \text{.}%
\end{equation}
\end{subequations}
In this case we find $\operatorname*{sgn}\lambda_{M\text{,}2}=\operatorname*{sgn}\lambda_{M\text{,}3}=\operatorname*{sgn}\alpha_{3}$ and $\operatorname*{sgn}\lambda_{M\text{,}4}=\operatorname*{sgn}\alpha_{1}$. $\lambda_{M\text{,}2}$ and $\lambda_{M\text{,}3}$ control the flows from/to $A_{3}$ and $N$, and $\lambda_{M\text{,}4}$ controls the flows from/to $A_{1}$ and $A_{2}$.

We are now in a position to give the directions of RG flows based on the local scaling exponents. The results are summarized below.

1. The flow between $N$ and $A_{3}$ is toward $N$ if $\alpha_{1}+\alpha_{2}>0$, and toward $A_{3}$ if $\alpha_{1}+\alpha_{2}<0$;

2. The flows between $N$ and $\chi^{\pm}$ are toward $N$ if $\alpha_{1}+\alpha_{2}+\alpha_{3}>0$, and toward $\chi^{\pm}$ if $\alpha_{1}+\alpha_{2}+\alpha_{3}<0$;

3. If $\alpha_{1}=\alpha_{2}$, the flows between $A_{3}$ and $A_{1,2}$ are toward $A_{3}$ if $\alpha_{1}<\alpha_{3}$, and toward $A_{1}$ and $A_{2}$ if $\alpha_{1}>\alpha_{3}$;

4. The flows between $A_{3}$ and $\chi^{\pm}$ are toward $A_{3}$ if $\alpha_{3}>0$, and toward $\chi^{\pm}$ if $\alpha_{3}<0$.

In addition, if the non-geometrical fixed point $M$ exists:

5. The flows between $M$ and $\chi^{\pm}$ are toward $M$ if $\lambda_{M\text{,}1}<0$, and toward $\chi^{\pm}$ if $\lambda_{M\text{,}1}>0$;

6. If $\alpha_{1}=\alpha_{2}$, the flows between $M$ and $A_{1,2}$ are toward $M$ if $\alpha_{1}<0$, and toward $A_{1}$ and $A_{2}$ if $\alpha_{1}>0$.

7. If $\alpha_{1}=\alpha_{2}$, the flow between $M$ and $A_{3}$ is toward $M$ if $\alpha_{3}<0$, and toward $A_{3}$ if $\alpha_{3}>0$.

8. If $\alpha_{1}=\alpha_{2}$, the flow between $M$ and $N$ is toward $M$ if $\alpha_{3}<0$, and toward $N$ if $\alpha_{3}>0$.

\subsection{RPA}

In the RPA, for a generic junction of $N$ quantum wires which is not on resonance, the RG equation for the S-matrix reads\cite{2016arXiv160100510S}

\begin{equation}
-\frac{dS_{jj^{\prime}}\left(  D\right)  }{d\ln D}=-\frac{1}{2}\sum
_{n_{1}n_{2}}\left[  S_{jn_{1}}\left(  D\right)  \Pi_{n_{1}n_{2}}\left(
D\right)  S_{n_{2}n_{1}}^{\ast}\left(  D\right)  S_{n_{2}j^{\prime}}\left(
D\right)  -\delta_{j^{\prime}n_{1}}\Pi_{n_{1}n_{2}}^{\ast}\left(  D\right)
S_{n_{2}n_{1}}\left(  D\right)  \delta_{n_{2}j}\right]  \text{.}
\label{RPANazarovGlazman}%
\end{equation}
where the vertex $\Pi\left(  D\right)  \equiv2\left[  \mathcal{Q}  -W\left(  D\right)  \right]  ^{-1}$, and the $D$ dependence of $W$ is through $S$. Again we suppress the $D$ dependence below.

\subsubsection{2-lead junction}

Calculating Eq.~(\ref{RPAvertex}) explicitly we find the cutoff-dependent RPA interaction

\begin{equation}
\Pi_{jj^{\prime}}=\frac{\left(  K_{1}^{-1}-1\right)  \left(  K_{2}%
^{-1}-1\right)  \left(  2\delta_{jj^{\prime}}-1\right)  W_{12}+2\left(
K_{j}^{-1}-1\right)  \delta_{jj^{\prime}}}{2+\left(  K_{1}^{-1}+K_{2}%
^{-1}-2\right)  W_{12}}\text{,} \label{2leadRPAvertex}%
\end{equation}
and the RPA RG equation for $S_{12}$,

\begin{equation}
-\frac{dS_{12}}{d\ln D}=-\frac{2\left(  1-W_{12}\right)  }{\gamma-1+2W_{12}%
}S_{12}\text{,}%
\end{equation}
where
\begin{equation}
\gamma=\frac{K_{1}^{-1}+K_{2}^{-1}+2}{K_{1}^{-1}+K_{2}^{-1}-2}\text{;}%
\end{equation}
or, in terms of $W_{12}$,

\begin{equation}
-\frac{dW_{12}}{d\ln D}=-\frac{4W_{12}\left(  1-W_{12}\right)  }%
{\gamma-1+2W_{12}}\text{,} \label{RPA2leadprob}%
\end{equation}
in agreement with the RPA RG equation for conductance in Ref.~\onlinecite{LithJPhys.52.2353}.

As with the first order equation Eq.~(\ref{Oalpha2leadprob}), Eq.~(\ref{RPA2leadprob}) has two fixed points, the $N$ fixed point $W_{12}=0$ and the $D$ fixed point $W_{12}=1$. In this case the $N$ fixed point has a conductance scaling exponent of $-4/\left(  \gamma-1\right)  =2-K_{1}^{-1}-K_{2}^{-1}$, and the $D$ fixed point has a scaling exponent of $4/\left(  \gamma+1\right)  =2-2\bar{K}$ where $\bar{K}=2/\left(  K_{1}^{-1}+K_{2}^{-1}\right)  $. Both exponents conform to the predictions of bosonic methods as has been verified in Ref.~\onlinecite{LithJPhys.52.2353}.

\subsubsection{Y-junction}

Starting from Eq.~(\ref{RPANazarovGlazman}) and following the same prescription which transforms Eq.~(\ref{NazarovGlazman}) to Eq.~(\ref{Oalpha3leadprob}), we find the RG equations obeyed by $a$, $b$, $c$ and $\bar{c}$ in the RPA:%

\begin{subequations}
\label{RPA3leadprob}%
\begin{align}
-\frac{da}{d\ln D}  &  =Q_{A}^{-1}\left\{  2\left[  \left(  1-b\right)
\left(  1+a+b-3a^{2}\right)  -\left(  1+3a\right)  c\bar{c}+c^{2}+\bar{c}%
^{2}\right]  \left(  Q_{1}+Q_{2}+Q_{3}-3\right)  \right. \nonumber\\
&  \left.  +\left(  3+a+3b-6a^{2}-ab-c\bar{c}\right)  \left(  Q_{1}%
+Q_{2}-2\right)  \left(  Q_{3}-1\right)  \right. \nonumber\\
&  \left.  +\sqrt{3}\left(  1-2a\right)  \left(  c+\bar{c}\right)  \left(
Q_{1}-Q_{2}\right)  \left(  Q_{3}-1\right)  +4\left[  a\left(  1-b\right)
-c\bar{c}\right]  \left(  Q_{1}-1\right)  \left(  Q_{2}-1\right)  \right\}
\text{,}%
\end{align}

\begin{align}
-\frac{db}{d\ln D}  &  =Q_{A}^{-1}\left\{  2\left[
\left(  1-b\right)  \left(  1+2b-3ab\right)  -3\left(  1+b\right)  c\bar
{c}\right]  \left(  Q_{1}+Q_{2}+Q_{3}-3\right)  \right. \nonumber\\
&  \left.  +\left[  \left(  1-b\right)  \left(  1+3a+2b\right)  -3c\bar
{c}\right]  \left(  Q_{1}+Q_{2}-2\right)  \left(  Q_{3}-1\right)  \right.
\nonumber\\
&  \left.  -\sqrt{3}\left(  1+2b\right)  \left(  c+\bar{c}\right)  \left(
Q_{1}-Q_{2}\right)  \left(  Q_{3}-1\right)  +4\left(  1-b\right)  \left(
1+2b\right)  \left(  Q_{1}-1\right)  \left(  Q_{2}-1\right)  \right\}
\text{,}%
\end{align}

\begin{align}
-\frac{dc}{d\ln D}  &  =Q_{A}^{-1}\left\{  2\left[
-c\left(  1+2b-3ab\right)  +\bar{c}\left(  2-2b-3c^{2}\right)  \right]
\left(  Q_{1}+Q_{2}+Q_{3}-3\right)  \right. \nonumber\\
&  \left.  +\left[  -c\left(  1+6a+2b\right)  +3\bar{c}\right]  \left(
Q_{1}+Q_{2}-2\right)  \left(  Q_{3}-1\right)  \right. \nonumber\\
&  \left.  +\sqrt{3}\left[  \left(  1+a\right)  \left(  1-b\right)
-2c^{2}-c\bar{c}\right]  \left(  Q_{1}-Q_{2}\right)  \left(  Q_{3}-1\right)
-4c\left(  1+2b\right)  \left(  Q_{1}-1\right)  \left(  Q_{2}-1\right)
\right\}  \text{,}%
\end{align}

\begin{align}
-\frac{d\bar{c}}{d\ln D}  &  =Q_{A}^{-1}\left\{  2\left[
-\bar{c}\left(  1+2b-3ab\right)  +c\left(  2-2b-3\bar{c}^{2}\right)  \right]
\left(  Q_{1}+Q_{2}+Q_{3}-3\right)  \right. \nonumber\\
&  \left.  +\left[  -\bar{c}\left(  1+6a+2b\right)  +3c\right]  \left(
Q_{1}+Q_{2}-2\right)  \left(  Q_{3}-1\right)  \right. \nonumber\\
&  \left.  +\sqrt{3}\left[  \left(  1+a\right)  \left(  1-b\right)  -2\bar
{c}^{2}-c\bar{c}\right]  \left(  Q_{1}-Q_{2}\right)  \left(  Q_{3}-1\right)
-4\bar{c}\left(  1+2b\right)  \left(  Q_{1}-1\right)  \left(  Q_{2}-1\right)
\right\}  \text{,}%
\end{align}
where%

\begin{align}
Q_{A}  &  =2\left[  \left(  1-a\right)  \left(
1-b\right)  -c\bar{c}\right]  \left(  Q_{1}+Q_{2}+Q_{3}-3\right)  +4\left(
1-b\right)  \left(  Q_{1}-1\right)  \left(  Q_{2}-1\right) \nonumber\\
&  +\left(  4-3a-b\right)  \left(  Q_{1}+Q_{2}-2\right)  \left(
Q_{3}-1\right)  -\sqrt{3}\left(  c+\bar{c}\right)  \left(  Q_{1}-Q_{2}\right)
\left(  Q_{3}-1\right) \nonumber\\
&  +6\left(  Q_{1}-1\right)  \left(  Q_{2}-1\right)  \left(  Q_{3}-1\right)
\text{.}%
\end{align}
\end{subequations}
In the special case of 1-2 symmetric interaction $K_{1}=K_{2}$, Eq.~(\ref{RPA3leadprob}) is reduced to the RG equations of Ref.~\onlinecite{PhysRevB.88.075131}. The fully 1-2 symmetric case with both $K_{1}=K_{2}$ and $c=-\bar{c}$ has been extensively analyzed there.

Eq.~(\ref{RPA3leadprob}) may once again be linearized to extract the scaling exponents. We first enumerate the scaling exponents at the geometrical fixed points $N$, $A_{j}$ and $\chi^{\pm}$; their physical meanings are identical to their first order counterparts.

$N$\textbf{ fixed point}: At $N$, $\lambda_{N\text{,}1}=2-K_{2}^{-1}-K_{3}^{-1}$, $\lambda_{N\text{,}2}=2-K_{3}^{-1}-K_{1}^{-1}$, $\lambda_{N\text{,}3}=2-K_{1}^{-1}-K_{2}^{-1}$, and $\lambda_{N\text{,}4}=3-K_{1}^{-1}-K_{2}^{-1}-K_{3}^{-1}$.

$A_{j}$\textbf{ fixed points}: At e.g. $A_{3}$, $\lambda_{A_{3}\text{,}1}=\lambda_{A_{3}\text{,}2}=2-K_{3}^{-1}-\left(  1+K_{1}K_{2}\right)  /\left(  K_{1}+K_{2}\right)  $, $\lambda_{A_{3}\text{,}3}=2-4K_{1}K_{2}/\left(  K_{1}+K_{2}\right)  $, and $\lambda_{A_{3}\text{,}4}=3-K_{3}^{-1}-\left(  1+3K_{1}K_{2}\right)  /\left(  K_{1}+K_{2}\right)  $.

$\chi^{\pm}$\textbf{ fixed points}: At $\chi^{\pm}$,

\begin{subequations}
\begin{equation}
\lambda_{\chi\text{,}j}=2-\frac{4\left(  K_{1}+K_{2}+K_{3}-K_{j}\right)
K_{j}}{K_{1}+K_{2}+K_{3}+K_{1}K_{2}K_{3}}\text{, }j=1\text{, }2\text{,
}3\text{;}%
\end{equation}

\begin{equation}
\lambda_{\chi\text{,}4}=3-\frac{4\left(  K_{1}K_{2}+K_{2}K_{3}+K_{3}%
K_{1}\right)  }{K_{1}+K_{2}+K_{3}+K_{1}K_{2}K_{3}}\text{.}%
\end{equation}
\end{subequations}

All scaling exponents above are in agreement with predictions of bosonization.\cite{PhysRevB.86.075451}

As for the non-geometrical fixed points, on account of mathematical simplicity we follow Ref.~\onlinecite{PhysRevB.88.075131} and only give their positions and scaling exponents in the fully 1-2 symmetric case, $K_{1}=K_{2}$ and $c_{0}=-\bar{c}_{0}$. We introduce the quantities

\begin{equation}
\zeta=\frac{3Q_{1}Q_{3}-Q_{1}-2Q_{3}}{2Q_{1}+Q_{3}-3}\text{,}%
\end{equation}

\begin{equation}
\tau_{0}=\sqrt{1+Q_{1}^{2}+2\zeta}\text{;}%
\end{equation}
where $Q_{1}$ and $Q_{3}$ are related to $K_{1}$ and $K_{3}$ by Eq.~(\ref{scriptQ}). $\zeta$ and $\tau_{0}$ are identical to $Q_{1}$ and $\tau$ in Ref.~\onlinecite{PhysRevB.88.075131} respectively.

$M$ \textbf{and} $Q$\textbf{ fixed points}: At these two fixed points%

\begin{equation}
\left(  a_{0},b_{0},c_{0}\right)  =\left(  \frac{1}{3}\left(  Q_{1}\mp\tau
_{0}\operatorname*{sgn}Q_{1}\right)  ,\frac{1}{6}\left(  \left(  \left\vert
Q_{1}\right\vert \mp\tau_{0}\right)  ^{2}-3\right)  ,0\right)  \text{,}
\label{abcMQ}%
\end{equation}
where the upper signs are for $M$ and the lower signs for $Q$. The two fixed points merge when $\tau_{0}=0$, or in terms of the Luttinger parameters,

\begin{equation}
K_{3}=\frac{2K_{1}\left(  K_{1}^{2}-K_{1}+1\right)  }{\left(  K_{1}+1\right)
\left(  2K_{1}-1\right)  }\text{.}%
\end{equation}
We note that $Q$ also exists for $Z_{3}$ symmetric interactions but its $W$ matrix remains $Z_{3}$ asymmetric; thus $Q$ cannot be reached when the RG flow starts from a $Z_{3}$ symmetric S-matrix. The $M$ fixed point again corresponds to the maximally open S-matrix in the $Z_{3}$ symmetric case, while the $Q$ fixed point only appears when the interactions are sufficiently strongly attractive.\cite{PhysRevB.84.155426,PhysRevB.88.075131} The conditions for $M$ and $Q$ to appear are $\tau_{0}^{2}=1+Q_{1}^{2}+2\zeta \geq0$ and $\left\vert \left\vert Q_{1}\right\vert \mp\tau_{0}\right\vert \leq3$ (the latter is due to S-matrix unitarity), and for $Z_{3}$ symmetric interactions $Q$ only starts to exist when the Luttinger parameter of all three wires $K\geq3$. Both $M$ and $Q$ are time-reversal symmetric.

For attractive interaction in wire $1$ ($K_{1}>1$), the $4$ scaling exponents at either $M$ or $Q$ are%

\begin{subequations}
\label{RPAMexps}%
\begin{equation}
\lambda_{M\left(  Q\right)  \text{,}1}=-\frac{3\left(  Q_{1}\pm\tau
_{0}+3\right)  \left(  Q_{1}^{2}+2Q_{1}-\tau_{0}^{2}\pm2\tau_{0}-3\right)
}{2\left(  Q_{1}\pm\tau_{0}\right)  \left(  2Q_{1}\mp\tau_{0}\right)  ^{2}%
}\text{,}%
\end{equation}

\begin{equation}
\lambda_{M\left(  Q\right)  \text{,}2}=-\frac{3\left(  \left(
Q_{1}\pm\tau_{0}\right)  ^{2}+3\right)  }{\left(  Q_{1}\pm\tau_{0}\right)
\left(  2Q_{1}\mp\tau_{0}\right)  }\text{,}%
\end{equation}

\begin{equation}
\lambda_{M\left(  Q\right)  \text{,}3}=-\frac{3\left(
Q_{1}\pm\tau_{0}-3\right)  \left(  Q_{1}^{2}-2Q_{1}-\tau_{0}^{2}\mp2\tau
_{0}-3\right)  }{2\left(  Q_{1}\pm\tau_{0}\right)  \left(  2Q_{1}\mp\tau
_{0}\right)  ^{2}}\text{,}%
\end{equation}

\begin{equation}
\lambda_{M\left(  Q\right)  \text{,}4}=\mp\frac{3\tau
_{0}\left(  \left(  Q_{1}\pm\tau_{0}\right)  ^{2}-9\right)  }{2\left(
Q_{1}\pm\tau_{0}\right)  \left(  2Q_{1}\mp\tau_{0}\right)  ^{2}}\text{;}%
\end{equation}
\end{subequations}
again the upper signs are for $M$ and the lower signs for $Q$. For repulsive interaction in wire $1$ ($K_{1}<1$), the lower signs should be taken to obtain the scaling exponents at $M$. When expanded to the first order in $\alpha_{1}$
and $\alpha_{3}$, Eq.~(\ref{RPAMexps}) agrees with Eq.~(\ref{OalphaMexps}).

$C^{\pm}$\textbf{ fixed points}: As with the $Q$ fixed point, the non-geometrical chiral fixed points $C^{\pm}$ only exist when the interaction is strongly attractive. At these fixed points

\begin{equation}
\left(  a_{0},b_{0},c_{0}\right)  =\left(  \frac{1}{6}\left(  2Q_{1}\left(
\zeta+2\right)  -\zeta^{2}+1\right)  ,\frac{1}{6}\left(  \zeta^{2}%
+4\zeta+1\right)  ,\pm\frac{\left(  \zeta-1\right)  }{6}\sqrt{3-\left(
2Q_{1}-\zeta\right)  \left(  2+\zeta\right)  }\right)  \text{.} \label{abcCpm}%
\end{equation}
The conditions for $C^{\pm}$ to appear are $-5\leq\zeta\leq1$ and

\begin{equation}
0\leq3-\left(  2Q_{1}-\zeta\right)  \left(  2+\zeta\right)  \leq\frac{1}%
{3}\left(  5+\zeta\right)  ^{2}\text{.}%
\end{equation}
In the $Z_{3}$ symmetric Y-junction, these conditions are satisfied when the Luttinger parameter of all three wires $K\geq2$. $C^{\pm}$ and $\chi^{\pm}$ merge when $\zeta=-2$, or in terms of the Luttinger parameters, $K_{1}^{-1}+2K_{3}^{-1}=1$.

There are again $4$ scaling exponents at $C^{\pm}$,

\begin{subequations}
\begin{equation}
\lambda_{C\text{,}1}=\frac{12Q_{1}}{\left(  Q_{1}-1\right)  \left(
3Q_{1}-\zeta-2\right)  }+\frac{12}{\left(  Q_{1}-1\right)  \left(
\zeta-1\right)  }\text{,}%
\end{equation}

\begin{equation}
\lambda_{C\text{,}2}=-\frac{12}{3Q_{1}-\zeta-2}-\frac{12}{\zeta-1}-6\text{,}%
\end{equation}

\begin{equation}
\lambda_{C\text{,}3\left(  4\right)  }=-\frac{3\left(  3Q_{1}-\zeta+4\right)
\left(  \zeta+5\right)  }{2\left(  3Q_{1}-\zeta-2\right)  \left(
\zeta-1\right)  }\left(  \frac{\zeta+1}{\zeta+5}\pm\frac{1}{3}\sqrt
{1+\frac{8\left(  \zeta+2\right)  \left(  3Q_{1}-\zeta-2\right)  }{\left(
\zeta-1\right)  \left(  3Q_{1}-\zeta+4\right)  }}\right)  \text{.}%
\end{equation}
\end{subequations}

We conclude this section with a discussion of the RG flows in the Y-junction.

In the generic $Z_{3}$ asymmetric case, for simplicity we only focus on the RG flows when $K_{1}$, $K_{2}$ and $K_{3}$\ are all close to unity, so that the only allowed non-geometrical fixed point is $M$. We focus on the flows between the geometrical fixed points. The results are listed below.

1. The flow between $N$ and $A_{3}$ is toward $N$ if $K_{1}^{-1}+K_{2}^{-1}>2$, and toward $A_{3}$ if $K_{1}^{-1}+K_{2}^{-1}<2$;

2. The flows between $N$ and $\chi^{\pm}$ are toward $N$ if $K_{1}^{-1}+K_{2}^{-1}+K_{3}^{-1}>3$ and $\lambda_{\chi\text{,}4}>0$, and toward $\chi^{\pm}$ if $K_{1}^{-1}+K_{2}^{-1}+K_{3}^{-1}<3$ and $\lambda_{\chi\text{,}4}<0$.

3. The flows between $A_{3}$ and $\chi^{\pm}$ are toward $A_{3}$ if $K_{3}^{-1}+\left(  1+K_{1}K_{2}\right)  /\left(  K_{1}+K_{2}\right)  >2$, and toward $\chi^{\pm}$ if $K_{3}^{-1}+\left(  1+K_{1}K_{2}\right)  /\left(  K_{1}+K_{2}\right)  <2$.

In the 1-2 symmetric case $K_{1}=K_{2}$, $W_{13}=W_{32}$ and $W_{23}=W_{31}$, Ref.~\onlinecite{PhysRevB.88.075131} has analyzed the stability of allowed fixed points based on the scaling exponents. The results are reproduced in Table~\ref{tab:RPA3leadFPstab}.

\begin{table}
\caption{Stability of 1-2 symmetric fixed points in the RPA found in Ref.~\onlinecite{PhysRevB.88.075131}, based on the RG phase portrait on the $\alpha_{1}$-$\alpha_{3}$ plane, Fig.~\ref{fig:3leadRPAphpo}. An unstable fixed point is denoted by a ``u'', a stable fixed point is denoted by an ``s'', and ``-'' indicates the corresponding fixed point does not exist in that region. The $M$ fixed point in region 12 is unstable against a 1-2 symmetry breaking perturbation, but is stable otherwise. Other fixed points have the same stabilities against 1-2 symmetry preserving and 1-2 symmetry breaking perturbations.\label{tab:RPA3leadFPstab}}
\begin{tabular}[c]{|l|lll|lll|}
\hline
& $N$ & $A_{3}$ & $\chi^{\pm}$ & $M$ & $Q$ & $C^{\pm}$\\ \hline
1 & u & s & u & - & - & -\\
2 & u & s & u & - & - & u\\
3 & u & u & s & u & - & -\\
4 & u & u & s & s & - & u\\
5 & u & s & u & s & u & -\\
6 & u & s & u & s & u & u\\
3' & u & u & u & s & - & -\\ \hline
7 & u & s & u & - & - & -\\
8 & u & s & u & u & - & -\\
9 & s & u & u & u & - & -\\
10 & s & u & u & u & - & -\\
11 & s & u & u & - & - & -\\
12 & u & u & u & s/u & - & -\\
\hline
\end{tabular}
\end{table}

Finally, following Ref.~\onlinecite{PhysRevB.88.075131} we detail the RG flows in a fully $Z_{3}$ symmetric junction with Luttinger parameter $K$ for all three wires. Only the fixed points consistent with $Z_{3}$ symmetry, namely $N$, $\chi^{\pm}$, $M$, and $C^{\pm}$, need to be considered.

When $0<K<1$, $N$ is the most stable fixed point, $M$ is stable against chiral perturbations but otherwise unstable, and $\chi^{\pm}$ are completely unstable; $C^{\pm}$ do not exist. The flows are from $\chi^{\pm}$ to $M$ or $N$, and from $M$ to $N$.

When $1<K<2$, $\chi^{\pm}$ are the most stable fixed points, $M$ is stable against time-reversal symmetric perturbations and unstable against chiral perturbations, and $N$ is completely unstable; $C^{\pm}$ do not exist. The flows are from $N$ to $M$ or $\chi^{\pm}$, and from $M$ to $\chi^{\pm}$.

When $2<K<3$, $\chi^{\pm}$ remain stable, $N$ remains completely unstable, while $M$ becomes fully stable. $C^{\pm}$ emerge as the unstable fixed points separating $\chi^{\pm}$ and $M$, approaching $\chi^{\pm}$ as $K$ approaches $3$. The flows are from $N$ to $M$, $\chi^{\pm}$ or $C^{\pm}$, and from $C^{\pm}$ to $M$ or $\chi^{\pm}$.

When $K>3$, $\chi^{\pm}$ become completely unstable, $N$ remains completely unstable and $M$ remains fully stable. $C^{\pm}$ remain unstable, moving toward $a_{0}=b_{0}=-1/3$ and $c_{0}=-\bar{c}_{0}=\pm2/3$ as $K\to \infty$. The flows are from $N$ to $M$ or $C^{\pm}$, from $\chi^{\pm}$ to $M$ or $C^{\pm}$, and from $C^{\pm}$ to $M$.

\bibliography{YJunctionentropy}

\end{document}